%% file: 00_paper.tex
\documentclass[sigconf, 10pt,nonacm]{acmart}
\usepackage[utf8]{inputenc}
\usepackage{xspace}
\usepackage{subfiles}
\usepackage{multirow}
\usepackage{tabularx}
\usepackage{subcaption}
\usepackage{enumitem}
\usepackage{balance}
\usepackage{xcolor}
\usepackage{cleveref}
\usepackage[splitrule,bottom]{footmisc}
\usepackage{threeparttable}
\usepackage{longtable}
\usepackage{graphicx} 
\usepackage{rotating} 
\usepackage{lipsum}  
\usepackage{float} 
\usepackage{placeins} 
\usepackage{array} 
\usepackage{soul}
\usepackage{makecell}
\usepackage[font=small, skip=0.5ex]{caption}
\usepackage{amsmath}
\usepackage{booktabs}
\usepackage{pifont}
\usepackage{longtable}
\usepackage{multirow}
\usepackage{array}
\usepackage{supertabular}
\usepackage{siunitx}
\usepackage[font=small, skip=0.5ex]{caption}

\usepackage{afterpage}
\usepackage{float}
\usepackage{booktabs,multirow,makecell,adjustbox}
\usepackage[subtle]{savetrees}

\newcolumntype{Y}{>{\centering\arraybackslash}X}

\newcommand{\sysguard}{\textit{GaitGuard}\xspace}
\newcommand{\sysextract}{\textit{GaitExtract}\xspace}

\newsavebox{\wideFigBox}

\renewcommand\footnotetextcopyrightpermission[1]{}
\setcopyright{none}

\AtBeginDocument{
  }

\begin{document}

\title{GaitGuard: Protecting Video-Based Gait Privacy \\in Mixed Reality}

\author{Diana Romero}
\email{dgromer1@uci.edu}
\affiliation{%
  \institution{University of California, Irvine}
  \city{}
  \state{}
  \country{}
}
\author{Athina Markopoulou}
\email{athina@uci.edu}
\affiliation{%
  \institution{University of California, Irvine}
  \city{}
  \state{}
  \country{}
}
\author{Salma Elmalaki}
\email{salma.elmalaki@uci.edu}
\affiliation{%
  \institution{University of California, Irvine}
  \city{}
  \state{}
  \country{}
}

\renewcommand{\shorttitle}{GaitGuard}

\renewcommand{\shortauthors}{Romero et al.}

\settopmatter{authorsperrow=3}

\begin{abstract}
Mixed Reality (MR) systems capture continuous video streams that expose bystanders' and collaborators' gait patterns—a biometric revealing sensitive attributes including age, gender, and health conditions. We show that video-based gait profiling achieves 78\% accuracy (15.6$\times$ random chance) on unprotected MR feeds, motivating \textbf{GaitGuard}, a real-time defense operating on a companion mobile device. GaitGuard introduces \textbf{GaitExtract}, an automated gait feature extraction pipeline adapted from clinical analysis for egocentric MR perspectives. Through systematic evaluation of 233 mitigation configurations, we characterize privacy-utility-performance trade-offs. A key insight is that gait features derive primarily from transient events (heel strikes, toe-offs). We exploit this temporal sparsity through adaptive mitigation that selectively processes only gait-critical frames, achieving a 68\% reduction in profiling accuracy while preserving visual quality (SSIM: 0.97) at 29~FPS. \textbf{GaitGuard} scales to 10 simultaneous users with under 10ms latency. A qualitative study of 20-participants confirms that the users preferred a solution such as \textbf{GaitGuard} which provides privacy guarantees. 

\end{abstract}



\maketitle

\input{sections/1-introduction}
\input{sections/2-background}

\input{sections/3-threat}
\input{sections/4-system}

\input{sections/5-evaluation}
\input{sections/6-discussion}
\input{sections/7-conclusion}

\begin{acks}
This research was partially supported by NSF awards 2339266 and 1956393, and a gift from the Noyce Initiative. We would like to thank Ruchi Patel for her contributions to the HoloLens implementation and data collection.
\end{acks}

\balance
\bibliographystyle{ACM-Reference-Format}
\bibliography{paper-base}
\newpage
\appendix
\input{sections/8-appendix}
\end{document}

%% file: sections/1-introduction.tex
\section{Introduction}\label{sec:introduction}

Mixed Reality (MR) has matured from research prototypes into mainstream consumer devices such as Apple Vision Pro~\cite{AppleVisionPro}, Meta Quest 3~\cite{MetaQuestNew}, and Meta Aria glasses~\cite{aria}. By blending physical and digital environments through advanced sensors for hand-tracking, eye-tracking, and speech input, these systems enable transformative applications in manufacturing~\cite{gonzalez-franco_immersive_2017}, healthcare~\cite{Prokopetc_2019_ICCV}, education~\cite{taherisadr2023erudite}, and remote collaboration~\cite{romero2025mocomr, cheng2024magicstream}.

This increasing adoption, however, raises significant privacy concerns. Sensor data and camera feeds can be exploited to extract sensitive information including facial data~\cite{corbettBystandARProtectingBystander2023}, semantic locations~\cite{farrukhLocInInferringSemantic2023}, and behavioral patterns~\cite{jarin2023behavr, chandio2020spatiotemporal}. These risks are amplified in multiuser MR scenarios, whether co-located in shared physical spaces or distributed across remote collaborations, affecting both primary users and bystanders.

\begin{figure}[!t]
\centering
\includegraphics[width=0.9\columnwidth, trim=40 100 60 210, clip]{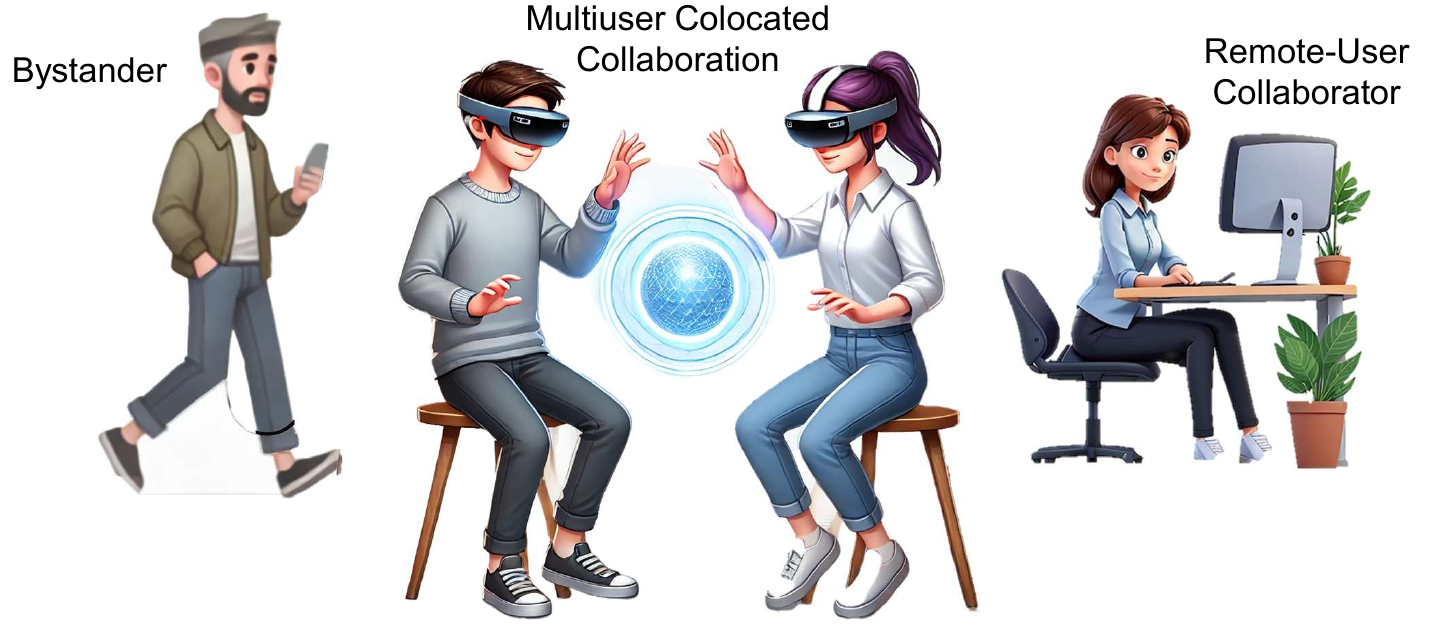}
\caption{Collaborative MR on a shared virtual object: a bystander walks into the headset's view while a remote desktop user assists, exposing moving users to potential video-based gait-profiling.}
\label{fig:teaser}
\end{figure}

This paper addresses a particularly underexplored privacy risk: gait profiling. Gait, the manner in which a person walks, is a recognized biometric used in medical diagnosis~\cite{baker2016gait} and mobile authentication~\cite{xu_gait-watch_2017}. Unlike other biometrics, gait patterns can reveal personal attributes that are not directly observable, including ethnicity~\cite{zhangEthnicityClassificationBased2010}, age~\cite{zhouDetectionAgeGroups2020}, gender~\cite{yuStudyGaitBasedGender2009}, and health conditions~\cite{QuantitativeGaitAnalysis}. Recognizing these risks, privacy regulations such as CPRA~\cite{CPRA} and GDPR~\cite{GeneralDataProtection} explicitly classify gait as protected biometric information.

MR devices continuously record their surroundings to enable core functionality, inadvertently capturing gait information in the process. Adversaries can exploit this through \textbf{gait profiling}: extracting distinctive gait features from video to infer personal information. As Figure~\ref{fig:teaser} illustrates, collaborative MR scenarios expose both primary users and bystanders to such attacks. Combined with machine learning, extracted gait features can reveal physical and mental health conditions, and predict future behavior~\cite{10.1145/3613905.3651073}, or even uniquely identify people~\cite{hanischUnderstandingPersonIdentification2023}, posing significant privacy risks.

We present \sysguard, a real-time system that protects gait privacy in MR environments. To that end, we develop system components and take intermediate steps, some of which are contributions on their own. In summary, the paper makes the following contributions.

\textbf{\sysguard~system:} We introduce the first real-time system designed to protect video-based gait profiling in collaborative MR. \sysguard has a modular architecture enabling the integration of new privacy mechanisms as they emerge. We provide a prototype implementation  ``locally'' on the device's companion app that mitigates data before external sharing to prevent downstream profiling.

\textbf{\sysextract:} We develop a fully automated pipeline that adapts clinical gait analysis to egocentric MR perspectives. This enables the systematic evaluation of profiling attacks and defenses without requiring manual calibration.

\textbf{Gait event dataset:} We construct the first MR dataset to explicitly label critical temporal transitions, comprising 109,200 frames across 1,197 walking sequences. Unlike prior works, we annotate the full temporal context (pre/post-event) of heel-strikes and toe-offs to train robust adaptive classifiers.

\textbf{Evaluation of mitigation strategies:} We rigorously evaluate 233 mitigation configurations, varying obfuscation regions (locations and sizes), noise types, and intensity levels. We characterize the resulting privacy-utility trade-offs to guide real-time deployment decisions.

\textbf{Adaptive mitigation:} We propose a learning-based strategy that selectively obfuscates infrequent but critical gait events. This approach significantly improves privacy while minimizing visual distortion compared to frame-agnostic obfuscation.

The paper proceeds as follows. Section 2 provides background and  Section 3 describes the threat model. Sections 4 and 5  present the \sysguard~design and the adaptive mitigation, respectively. Section 6 presents the implementation and Section 7 present the evaluation. Section 8 discusses future work and Section 9 concludes the paper.

%% file: sections/2-background.tex
\section{Background and Related Work}\label{sec:background}

\subsection{Gait Analysis Methods}\label{sec:gait-analysis}

\begin{figure}[!t]
  \centering
  \includegraphics[width=0.9\linewidth, trim=70 130 80 234, clip]{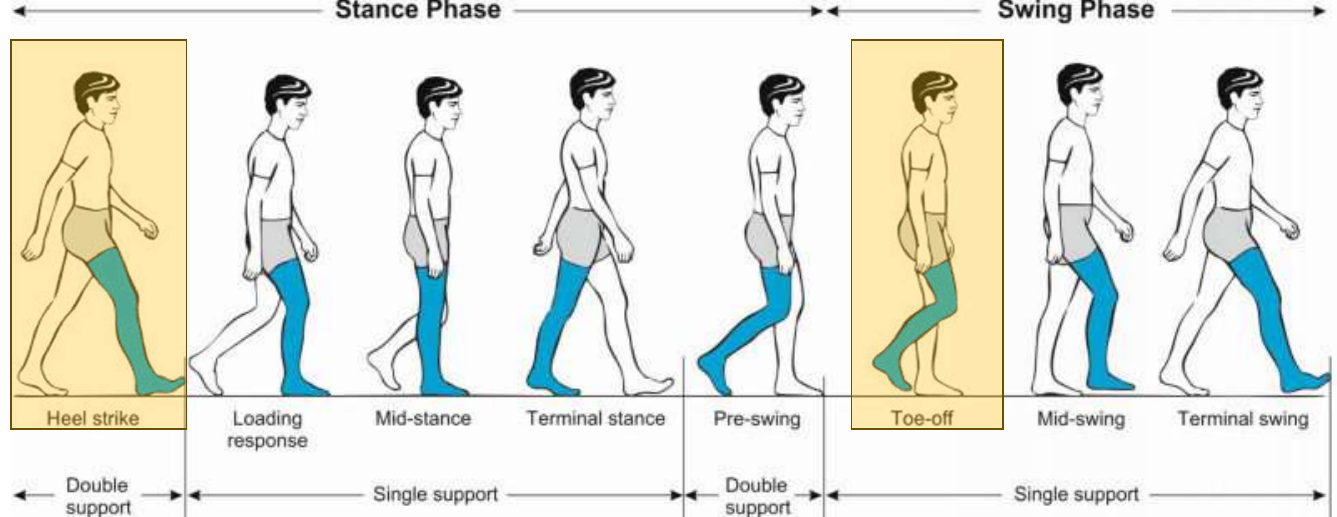}
  \caption{Gait cycle highlighting toe-off and heel strike events~\cite{pirker2017gait}.}
  \label{fig:gait-cycle}
\end{figure}

Gait analysis is the systematic study of human locomotion. It characterizes walking as a series of gait cycles, where each cycle (or stride) spans the period between successive ground contacts of the same foot. Within each cycle, critical \textit{gait events} occur, including \textit{heel strike} and \textit{toe-off}, as illustrated~in~Figure~\ref{fig:gait-cycle}. These events form the basis for extracting \textit{gait features}, quantitative metrics describing the temporal and spatial properties of walking~\cite{stenumTwodimensionalVideobasedAnalysis2021}, including:

\begin{itemize}[leftmargin=0.3cm, noitemsep, topsep=0cm]
    \small
    \item \emph{Step time (L/R)}: Seconds between consecutive bilateral heel strikes.
    \item \emph{Stance time (L/R)}: Seconds from heel strike to toe-off per leg.
    \item \emph{Swing time (L/R)}: Seconds from toe-off to heel strike per leg.
    \item \emph{Double support time}: Seconds when both feet contact the ground.
    \item \emph{Step length (L/R)}: Distance between ankles at heel strike.
\end{itemize}

Instrumented gait analysis (IGA) using motion capture systems, force plates, and instrumented treadmills represents the gold standard in clinical and research settings~\cite{cappozzoGaitAnalysisMethodology1984}. However, the high cost and laboratory constraints of IGA have prompted the exploration of alternative sensing modalities, including inertial measurement units (IMUs)~\cite{prasanthWearableSensorBasedRealTime2021}, gyroscopes~\cite{ReliableGaitPhase}, and depth sensors such as Microsoft Kinect~\cite{gabelFullBodyGait2012}.

More recently, video-based gait analysis using 2D pose estimation has emerged as a compelling alternative, with widespread algorithms like OpenPose demonstrating strong agreement with IGA measurements(Pearson $r > 0.9$) by tracking body keypoints over time~\cite{stenumTwodimensionalVideobasedAnalysis2021, caoOpenPoseRealtimeMultiPerson2019}. This progression from clinical IGA to video-based methods is significant: it means that gait profiling no longer requires specialized equipment. Any camera, including those embedded in MR headsets, can potentially extract clinically meaningful gait features. In this paper, we focus on this threat vector: video-based gait profiling from MR-captured camera streams.

\subsection{Gait Privacy Beyond Identification}\label{sec:gait-privacy}

Gait is a well-known biometric widely studied for authentication~\cite{colaGaitbasedAuthenticationUsing2016, delgado-santosGaitPrivacyONPrivacypreservingMobile2022a} and, unlike passwords, it cannot be easily concealed or changed~\cite{hanischUnderstandingPersonIdentification2023}, making it an attractive target for surveillance. Gait reveals far more than identity. It encodes sensitive attributes including age, gender, ethnicity, and physical or neurological health~\cite{zhangEthnicityClassificationBased2010, zhouDetectionAgeGroups2020, yuStudyGaitBasedGender2009, QuantitativeGaitAnalysis}. Deviations in gait patterns can indicate early-stage neurodegenerative diseases such as Parkinson's~\cite{mirelman2019gait} or signal cognitive decline~\cite{beauchet2016poor}. Unlike facial features or explicit identifiers, these inferences can be drawn unobtrusively, making gait uniquely revealing and difficult to protect. A notable example is actor Billy Connolly, whose Parkinson's disease was diagnosed after a medical professional noticed a slight gait alteration, underscoring the intimate information contained in walking patterns~\cite{Connolly2014}.

While several efforts have explored privacy-preserving methods for wearables—such as IMU perturbation~\cite{rasnayaka_your_2020} or sensor obfuscation~\cite{xia_privacy-aware_2023}—these approaches typically assume direct access to user-owned devices.

MR presents a distinct challenge: onboard cameras passively capture gait data from users and unaware bystanders without instrumentation. We focus on \textbf{gait profiling}, extracting gait features from video to infer personal information. Despite the sensitive nature of this data and the ubiquity of head-mounted cameras in MR, no prior work has systematically addressed gait privacy in this context.

\subsection{Datasets for Gait Analysis}
Recent contributions to the gait data collection~\cite{nature_gait_collection}, such as those by Luo et al.~\cite{Luo2020} and Losing et al.~\cite{Losing2022}, prioritize kinematic data from wearables in naturalistic settings, often deriving events like heel-strike from sensor signals rather than direct visual observation. While video-based datasets like Health\&Gait~\cite{ZafraPalma2025} address visual analysis, they predominantly target aggregate parameters or pose estimation, lacking granular temporal annotations of specific gait phases. To address this gap, we construct a labeled gait event dataset from MR-captured footage containing 109,200 datapoints across 1,197 walking sequences. Unlike prior works, our strategy explicitly labels frames immediately before, during, and after heel-strike and toe-off events to capture the full temporal context required for adaptive mitigation.

\subsection{Privacy in MR}\label{sec:mr-privacy}
MR devices capture extensive video, audio, and motion information, posing risks not only to device users but also to non-consenting bystanders. Prior work has documented diverse vulnerabilities in MR environments, ranging from visual data leaks~\cite{10.1145/3325424.3329659} and perceptual manipulation~\cite{chengExploringUserReactions2023} to sensor-based side-channels such as keystroke inference~\cite{zhang2024vreckey} and voice reconstruction~\cite{li2024speakup}. These findings underscore the potential for sensitive behavioral and biometric data leakage in immersive systems. While efforts such as BystandAR~\cite{corbettBystandARProtectingBystander2023} have addressed facial privacy protection, gait privacy in MR environments remains largely unexplored.

\begin{figure*}[!t]
  \centering
  \includegraphics[width=0.99\linewidth, trim=0 205 15 180, clip]{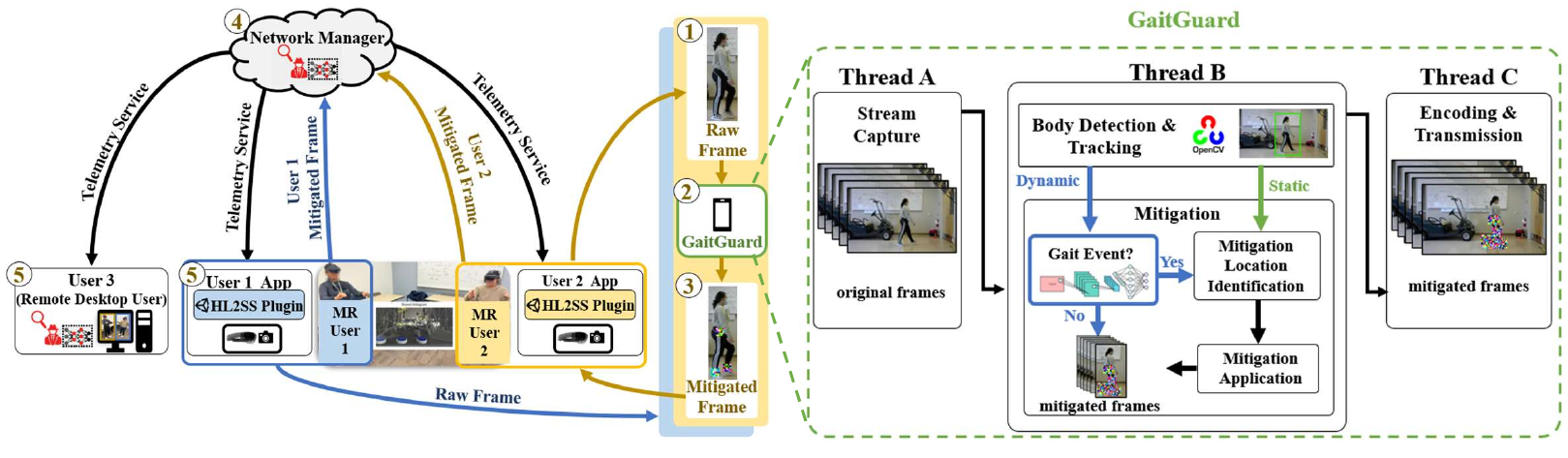}
  \caption{\textbf{Overview of \sysguard.} \textbf{[Left]} Multiple users interact in a collaborative MR app. \sysguard: \ding{182}~intercepts raw frames from HoloLens~2; \ding{183}~\sysguard applies gait mitigation; \ding{184}~releases mitigated frames to the app—protecting against adversaries on the \ding{185}~server or \ding{186}~remote clients. \textbf{[Right]} \sysguard implemented as three pipelined threads: \textbf{A}~captures frames, \textbf{B}~detects humans and mitigates gait, \textbf{C}~transmits output—maintaining real-time MR performance. An \textit{honest-but-curious} adversary may reside in the telemetry manager or remote desktop client. }
  \label{fig:sys-overview}
  
\end{figure*}

\textbf{Why Gait Privacy Matters.} While one might argue that facial recognition poses a more direct identification threat, gait privacy remains critical for several reasons. First, when faces are obscured by systems like BystandAR, gait remains one of the few \emph{persistent} biometric vectors—unlike clothing or accessories, it cannot be changed between sessions, making it valuable for longitudinal tracking. Second, gait recognition operates where face recognition fails: it functions at distances exceeding 50 meters, from any viewing angle, and at resolutions too low for reliable facial identification~\cite{nixon2007automatic}. Third, gait is explicitly protected as biometric data under GDPR Article~9, CCPA/CPRA, and Illinois BIPA~\cite{GeneralDataProtection, CPRA}, creating independent compliance obligations. Fourth, as discussed in \S\ref{sec:gait-privacy}, gait reveals health conditions that facial features do not. Finally, defense-in-depth principles suggest protecting multiple biometric vectors: even when faces are captured, denying gait data reduces the attack surface and prevents cross-modal correlation attacks.

%% file: sections/3-threat.tex
\section{Threat Model}\label{sec:threat}

Any device with access to camera streams containing walking individuals can extract gait features. While this threat applies broadly to camera-enabled devices, collaborative MR environments amplify the risk: camera feeds are continuously transmitted for shared perception, exposing multiple users and bystanders to profiling. Our analysis of $318$ HoloLens applications found that $26\%$ request camera access.

\textbf{Attack Scenarios.}
Gait profiling can happen in realistic MR deployments: (1)~enterprise IT administrators accessing video logs from tools like Microsoft Dynamics 365 Remote Assist~\cite{microsoft_dynamics_remote_assist_overview} for ``quality assurance''; (2)~compromised relay infrastructure where honest-but-curious operators at services like Photon~\cite{PhotonUnityNetworking} mine camera feeds from concurrent sessions; (3)~malicious remote collaborators who record incoming video to profile participants; and (4)~legal discovery processes that expose bystander gait from subpoenaed MR recordings.

\textbf{Scope.} We focus on gait, rather than facial privacy, because gait presents a distinct, complementary threat as discussed in \S\ref{sec:mr-privacy}: it persists when faces are obscured, functions at greater distances, and reveals health information. \sysguard can operate alongside face-protection systems like BystandAR to provide defense-in-depth.

\textbf{Attacker Model.}
As depicted in Figure~\ref{fig:sys-overview}, we consider two primary attackers:
\emph{Attacker 1: Central Service Operator}---has privileged access to data passing through relay infrastructure and can extract gait from multiple users' streams.
\emph{Attacker 2: Remote Collaborating User}---receives shared camera streams and can passively analyze them to infer gait patterns.

\textbf{Attacker Goal.}
The attacker extracts distinctive gait features from RGB frames to infer personal information about users or bystanders, including identity, health conditions, age, and gender.

\textbf{Proposed Defense.}
We propose \sysguard, a companion application on a trusted and paired mobile device that intercepts video frames and applies privacy-preserving transformations before transmission. By protecting streams before network transmission, \sysguard mitigates risks from both attacker types while maintaining real-time MR performance. Although we focus on MR, our approach can be applied to  protect gait privacy in any video capture.

%% file: sections/4-system.tex
\section{\sysguard System Design}\label{sec:system}

\sysguard mitigates gait features while preserving visual quality by processing MR headset video frames on a paired mobile device, as illustrated in Figure~\ref{fig:sys-overview}. The system balances privacy protection with real-time performance through a modular architecture that can accommodate new mitigation techniques as they emerge. Two constraints guide our design: (1)~video transformations must preserve content required for MR functionalities such as spatial mapping and avatar rendering, and (2)~commercial MR headsets have restricted platforms that prevent on-device camera stream interception.

\subsection{Architecture and Deployment Strategy}

\sysguard offloads privacy processing to a companion mobile device. Camera frames from the MR headset are streamed in real time to the mobile device, where gait detection and mitigation are applied. The transformed frames are then returned to the MR application pipeline. This design enables low-latency operation while offering broad compatibility with existing MR systems, allowing deployment without modification to headset firmware or applications. Importantly, the companion device does not require active user interaction during MR sessions—it can remain in a pocket or bag while communicating with the headset over Wi-Fi Direct or Bluetooth, imposing minimal burden on the user experience. This architecture is necessitated by platform limitations in commercial headsets like HoloLens~2, which prevent developers from intercepting or modifying raw camera streams~\cite{lolambeanHoloLensHardware2023}. Similar constraints have been encountered in prior MR privacy work such as BystandAR~\cite{corbettBystandARProtectingBystander2023}. Moreover, the computational cost of video-based gait analysis is prohibitive on headset hardware: OpenPose achieves 11 fps on a high-end Nvidia V100 GPU and less than 1 fps on standard CPUs~\cite{caoOpenPoseRealtimeMultiPerson2019}, well below the 30 fps needed for MR capture~\cite{vtietoMixedRealityCapture2022}.

\textbf{Architecture Alternatives.}
We considered several alternative architectures before selecting companion-device processing. \emph{Cloud processing} would introduce unacceptable latency and privacy risks, while \emph{edge server processing} requires dedicated infrastructure that limits deployability. \emph{On-headset processing} is infeasible given current platform restrictions and compute limitations discussed above. The companion smartphone represents an optimal balance: users already carry phones, modern devices (circa 2019+) provide sufficient compute for real-time inference, and processing remains under user control. As MR hardware evolves to include dedicated neural processing units—such as Apple Vision Pro's Neural Engine or Qualcomm's Snapdragon XR2 Gen 2—\sysguard's architecture can migrate to fully on-device processing.

\subsection{GaitExtract}\label{sec:gaitextract}
A key challenge in studying video-based gait privacy is the lack of tools for extracting clinical gait features from unconstrained video. While extensive work exists on gait recognition for identification~\cite{yu2006framework} (e.g., using silhouette-based methods on datasets like CASIA-B), these approaches classify identity rather than extract the temporal and spatial gait parameters (step time, stance time, swing time) that reveal sensitive health attributes. Methods that do extract such clinical features, such as the approach of Stenum et~al.~\cite{stenumTwodimensionalVideobasedAnalysis2021}, were designed for controlled laboratory settings and require manual intervention.

To address this gap, we developed \sysextract, an automated gait profiling tool that adapts the clinically validated methodology of Stenum et~al. for real-time use on unconstrained video. Critically, Stenum et~al. validated their approach against gold-standard instrumented gait analysis (IGA) using 3D motion capture systems, demonstrating strong agreement for temporal gait parameters (Pearson $r > 0.9$ for step time, stance time, and swing time)~\cite{stenumTwodimensionalVideobasedAnalysis2021}. Since both laboratory video and MR-captured video are fundamentally RGB image sequences analyzed through the same pose estimation pipeline, this validation transfers directly to our deployment context. \sysextract represents, to our knowledge, the first fully automated pipeline for extracting clinical gait features from video captured in uncontrolled environments.

\begin{figure}[!t]
    \centering
    \includegraphics[width=\linewidth, trim=0 90 0 90, clip]{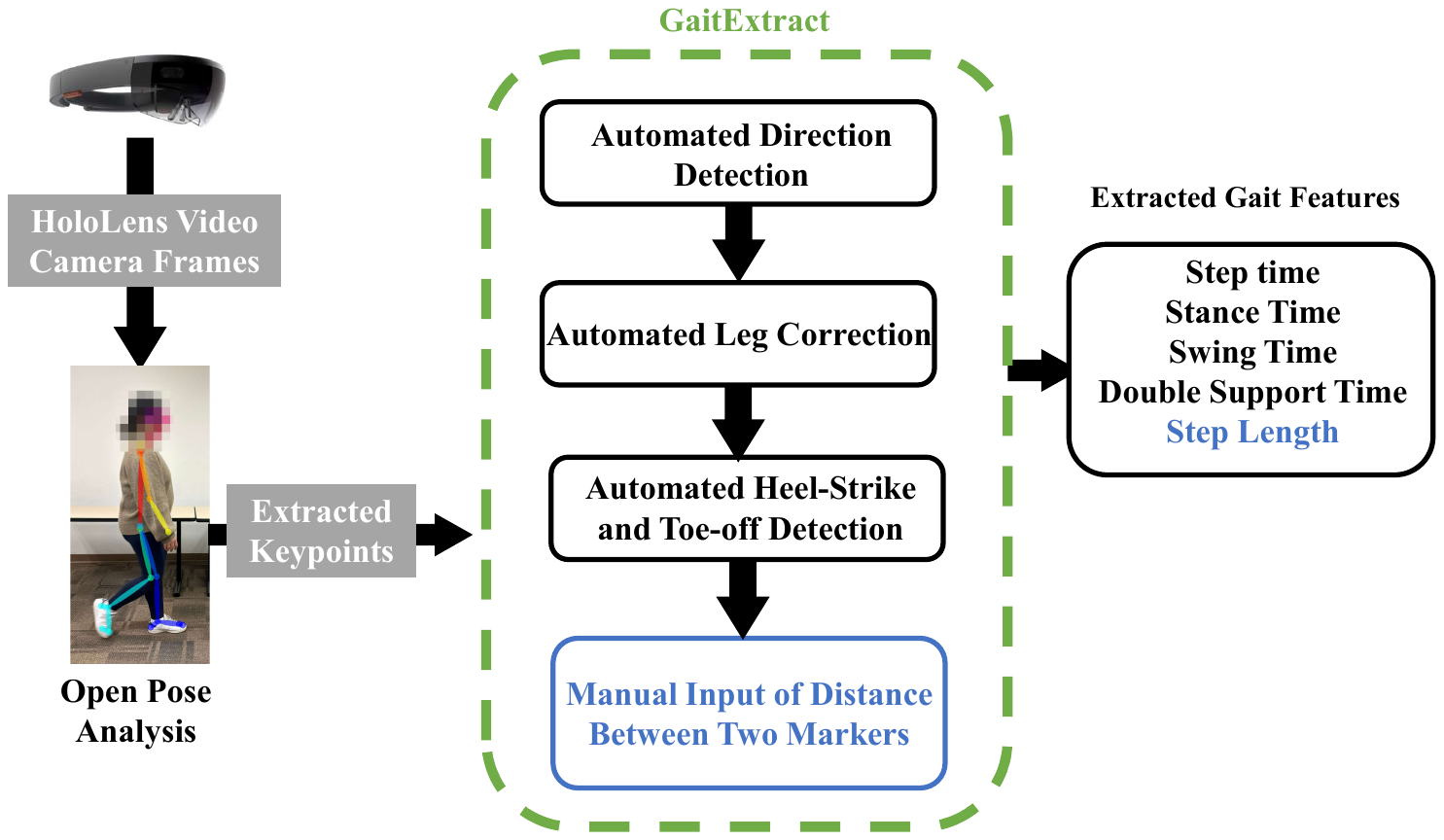}
    \caption{\sysextract: A gait-profiling pipeline using OpenPose to extract temporal features and estimate step length with minimal calibration.}
    \label{fig:gait-framework}
\end{figure}

As illustrated in Figure~\ref{fig:gait-framework}, \sysextract~advances video-based gait analysis by resolving the specific complexities of moving, egocentric cameras compared to fixed laboratory setups. Its contributions center on a robust pipeline that automates \emph{walking direction estimation} to handle dynamic viewpoints, while simultaneously deploying a \emph{leg-side correction} mechanism to rectify keypoint misclassifications caused by occlusion or noise. Beyond basic tracking, the system enables precise \emph{gait event detection}—identifying heel-strikes and toe-offs in unconstrained sequences—and ensures data quality through \emph{cycle filtering} that discards incomplete or interrupted walks. Finally, \sysextract~streamlines the analytical workflow via \emph{automated logging}, aggregating only valid features to provide a reliable basis for downstream profiling and defense evaluation.

\sysextract supports fully automated extraction of temporal gait features (step time, swing time, stance time, double support time) and a partially automated mode that additionally estimates step length with minimal calibration. For each walking sequence, the tool outputs a ten-dimensional feature vector per leg. These features, derived from a clinically validated approach, serve as ground truth for evaluating our mitigation strategies.

\subsection{Mitigation Design Space}
\label{sec:mitigation}

\begin{figure*}[!t]
  \centering
  \includegraphics[width=0.92\linewidth, trim=5 170 5 120, clip]{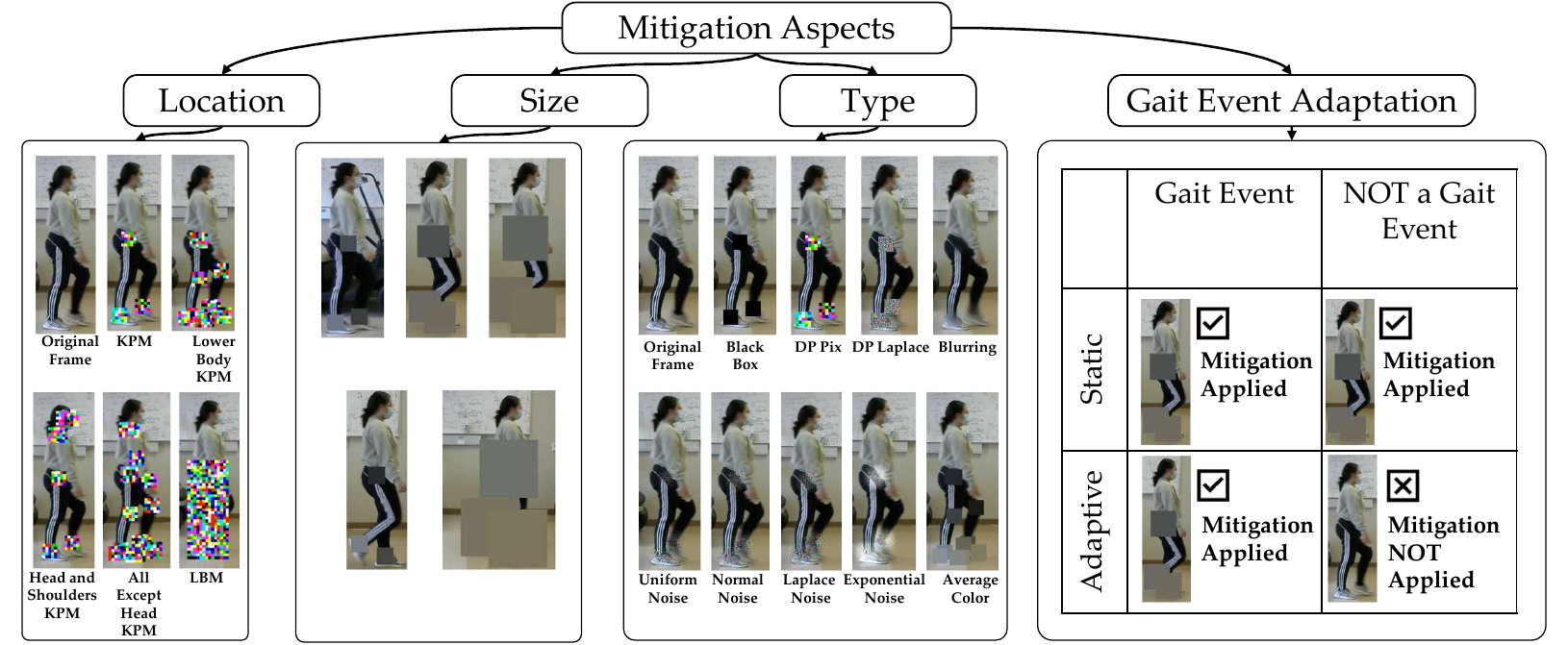}
  \caption{\textbf{Design space of gait mitigation strategies.} Mitigation techniques are explored along four aspects: \ding{182}~location of application, \ding{183}~mask size, \ding{184}~type of transformation, and \ding{185}~adaptability. Each axis presents trade-offs in privacy protection, visual quality, and computational efficiency.}
  \label{fig:mitigation}
  
\end{figure*}

We developed mitigation strategies spanning four design aspects: \ding{182}~\textbf{location}, \ding{183}~\textbf{size}, \ding{184}~\textbf{type}, and \ding{185}~\textbf{adaptability}, as summarized in Figure~\ref{fig:mitigation}. Our modular architecture allows these components to be configured independently, and new mitigation techniques can be integrated as they emerge.

\noindent\ding{182}~\textbf{Location: Where to Apply Mitigation.}
Gait features are primarily derived from lower body motion, particularly the timing and position of heel strikes and toe-offs at the ankles and hips~\cite{stenumTwodimensionalVideobasedAnalysis2021}. This suggests that targeted masking of gait-critical regions can protect privacy while preserving visual content elsewhere in the frame. We explore two spatial targeting approaches: Keypoint-Based Masking (KPM), which masks regions around specific anatomical landmarks detected via pose estimation, and Lower Body-Based Masking (LBM), which masks the entire lower body region using a bounding box.

Keypoint-Based Masking (KPM) applies transformations to regions surrounding body keypoints identified by OpenPose. We evaluated several configurations:

(1)~\emph{KPM baseline} masks three core gait-related keypoints (midhip, left ankle, right ankle) to directly disrupt heel strike and toe-off detection.

(2)~\emph{Head and shoulders KPM} adds upper body joints to explore whether concealing upper body movement enhances privacy.

(3)~\emph{Lower body KPM} focuses on lower body joints including hips, knees, heels, and toes.

(4)~\emph{All-except-head KPM} masks all body keypoints except the head, offering maximal gait obfuscation while retaining facial context.

Lower Body-Based Masking (LBM) implements a region-based strategy that masks the entire lower body using a bounding box. This approach avoids per-keypoint computation, reducing processing overhead while maintaining coverage of gait-critical areas~\cite{mancini2010relevance}.

\noindent\ding{183}~\textbf{Size: Area of Masking.}
Mask size balances privacy protection against visual quality. Larger masks provide greater concealment but introduce more distortion. For KPM, we evaluated box sizes of 50, 100, 150, and 200 pixels around each keypoint. LBM uses a fixed bounding box encompassing the lower body region.

\noindent\ding{184}~\textbf{Type: How to Obscure Gait Features.}
Various privacy-preserving transformations have been proposed, including adversarial perturbations~\cite{zajac2019adversarial}, GAN-based anonymization~\cite{sirichotedumrong2021gan}, and imperceptible filters~\cite{shen_human-imperceptible_2019}. However, many are computationally intensive and unsuitable for real-time applications~\cite{guesmi2022room}. We evaluate lightweight techniques that meet real-time constraints. Our modular design allows additional techniques to be incorporated as more efficient methods become available.

(1) \emph{Black Box} places a solid black rectangle over gait-relevant regions. This baseline technique guarantees concealment but introduces high visual distortion.

(2) \emph{Differentially Private Pixelization (DP-Pix)} adapts Fan et al.~\cite{fanImagePixelizationDifferential2018} by pixelizing regions and adding Laplace noise. We extended this to RGB by applying noise to each color channel, testing kernel sizes $b \in \{10, 20, 30, 40, 50\}$ with $\epsilon = 0.1$.

(3) \emph{Differentially Private Laplacian Noise (DP-Laplace)} applies Laplace noise directly to pixels without pixelization, testing $\epsilon \in \{0.1, 10, 100, 1000\}$.

(4) \emph{Gaussian Blur} applies blurring with kernel sizes $c \in \{5, 25, 45, 65\}$, similar to YouTube's face-blurring feature~\cite{noauthor_blur_nodate}.

(5) \emph{ Random Noise} adds noise from uniform, normal, Laplace, or exponential distributions with intensity parameter $\lambda \in \{50, 100, 150, 200\}$, using a Hanning window for smooth edges.

(6) \emph{Average Color} replaces masked regions with their mean RGB color, maintaining scene context while obscuring gait-relevant detail.

\textbf{Mitigation Combinations.}
Across location (\ding{182}), size (\ding{183}), and type (\ding{184}), we evaluate \emph{233 unique configurations}: 109 DP-Laplace, 56 DP-Pix, 8 Gaussian blur, 32 additive noise, and 28 average color and black-box settings.

\section{Adaptive Mitigation \& Gait Event Dataset}\label{sec:adaptive}

The mitigation strategies described in \S~\ref{sec:mitigation} are applied uniformly to all video frames, a mode we define as \emph{static mitigation}. Although effective, this approach can be resource-intensive and degrade both visual quality and processing performance. To address this, we explore the dimension of \textbf{adaptability}~(\ding{185}) by introducing an \emph{adaptive mitigation} strategy that selectively applies transformations only to frames containing gait-relevant information.

Adaptation could be based on various criteria: scene content, user proximity, or application context. We focus on adapting based on \emph{gait events} because gait features are fundamentally derived from the timing of specific biomechanical events within the gait cycle~\cite{pirker2017gait}. Specifically, most clinical gait parameters (step time, stance time, swing time) depend on accurately detecting heel-strike and toe-off events. By applying mitigation only to frames where these events occur, we can preserve privacy while minimizing unnecessary visual distortion during other phases of the walking cycle.

\textbf{Gait Event Detection.}
Based on gait cycle biomechanics~\cite{pirker2017gait}, heel-strike and toe-off are the most informative events for gait profiling, as they define the boundaries of stance and swing phases from which temporal features are computed. We define frames containing these events as ``gait event frames'' and propose applying mitigation selectively to these frames while leaving others unmodified.

\textbf{Gait Event Dataset.}
Training a gait event classifier requires labeled data distinguishing event frames from non-events. Existing datasets use motion capture~\cite{cappozzoGaitAnalysisMethodology1984}, IMUs~\cite{prasanthWearableSensorBasedRealTime2021}, or stationary side-view video~\cite{stenumTwodimensionalVideobasedAnalysis2021}. None addresses egocentric capture with mobile observers, variable distances, and partial occlusion. We created a dataset of 109,200 images from 1,197 walking sequences of 20 users recorded using HoloLens~2.

We developed our labeling methodology by adapting the clinically validated approach of Stenum et al.~\cite{stenumTwodimensionalVideobasedAnalysis2021} for identifying heel-strike and toe-off in videos. Our labeling strategy includes frames immediately before, during, and after heel-strike or toe-off events to capture the full temporal context of gait transitions. The dataset focuses on the lower body region, identified using YOLOv5~\cite{yolov5} for person detection and cropping the lower half of the bounding box. Our dataset reflects the natural distribution of gait and non-gait event frames across all collected data, providing a foundational resource for training gait event classification models.

\textbf{Gait Event Classifier.}
To enable real-time adaptive mitigation, we designed a convolutional neural network (ConvNet) that classifies whether a frame contains a gait event. Our goal was to develop an effective, lightweight classifier suitable for real-time MR applications rather than to achieve state-of-the-art classification accuracy.

We initially explored a two-stream architecture with separate pathways for spatial (RGB) and temporal (optical flow) learning. However, real-time optical flow computation introduced significant latency that compromised streaming performance. We therefore optimized to a single spatial ConvNet, illustrated in Appendix~\ref{app:convnet}, 
comprising five convolutional layers, two fully connected layers, and an output layer representing gait-event probability.

The network uses a 50\% probability threshold, selected to maximize recall and ensure all gait events are detected even at the risk of some false positives. To address class imbalance (gait events are a small fraction of frames), we employ Focal Loss with $\alpha = 0.8$ and $\gamma = 2.5$, achieving 94.3\% recall in gait event classification.

\section{Implementation}\label{sec:implementation}

\sysguard is deployed as a companion application written in Python, running alongside an MR application on HoloLens~2. The system consists of a multi-threaded, pipelined architecture, with each processing stage handled by a dedicated thread to ensure high throughput and real-time performance, as illustrated in Figure~\ref{fig:sys-overview}.

\textbf{Companion Application.}
The companion application pairs with the Microsoft Multi-user Mars Rover Application~\cite{jessemccullochIntroductionMultiuserCapabilities2022}, the baseline framework for multi-user applications on HoloLens~2. The application streams camera frames from HoloLens~2 to the \sysguard server, enabling integration with existing and future multi-user MR applications with minimal modification.

\textbf{Pipelined Thread Architecture.}
\sysguard is designed around a pipelined architecture with three threads: \textit{Thread~A} handles stream capture, \textit{Thread~B} handles body detection and mitigation, and \textit{Thread~C} handles encoding and transmission. These threads communicate via shared queues, allowing each stage to operate independently and asynchronously.

(1) \emph{Stream Capture (Thread A).}
This module receives raw video frames from the MR headset using HL2SS APIs to establish a connection and configure stream parameters including resolution, frame rate, and codec. Once connected, it continuously captures frames and adds them to an input queue. Operating in its own thread, this module is isolated from processing delays in downstream modules, enabling uninterrupted frame acquisition.

(2) \emph{Body Detection and Mitigation (Thread B).}
This thread processes incoming frames to detect humans and apply privacy-preserving transformations. Human body detection uses OpenCV's Histogram of Oriented Gradients (HOG) descriptor with a pretrained SVM classifier~\cite{OpenCVCvHOGDescriptor}. To reduce processing time, frames are resized to 320$\times$240 and converted to grayscale.

For gait mitigation, this thread applies the strategies described in \S~\ref{sec:mitigation}. In KPM, OpenPose locates keypoints (hips, ankles, knees) and targeted mitigation is applied to these regions. In LBM, the lower half of detected bounding boxes is masked. Mitigation is applied statically or adaptively as described in \S~\ref{sec:adaptive}. Processed frames are placed in an output queue for encoding.

(3) \emph{Encoding and Transmission (Thread C).}
This module retrieves mitigated frames, compresses them, and transmits them to the MR system. To reduce bandwidth, frames are resized to 30\% of original resolution and encoded using JPEG with quality setting 10. Transmission uses a UDP socket configured for high-throughput streaming. Running in a dedicated thread, this module ensures encoding and network operations do not interfere with upstream processing.

By structuring \sysguard as a multithreaded pipelined system, we maintain steady frame rates and minimize latency across capture, mitigation, and transmission stages. Critically, mitigation occurs on frames within the network pipeline without affecting the user's egocentric view or virtual objects.

%% file: sections/5-evaluation.tex
\section{Evaluation}\label{sec:evaluation}

We evaluated \sysguard's capability to reduce gait profiling while preserving real-time and visual quality in collaborative MR applications. Our study tested 233 strategies in static and adaptive modes (Section~\ref{sec:mitigation}), assessing privacy, utility, and system performance.

\subsection{Methodology}

We implemented \sysguard as a companion system for a collaborative MR application on HoloLens~2, with the mitigation module running on a Samsung S23 Ultra Android phone.

\begin{figure}[!t]
  \centering
    \includegraphics[width=0.7\linewidth, trim=0 90 0 90, clip]{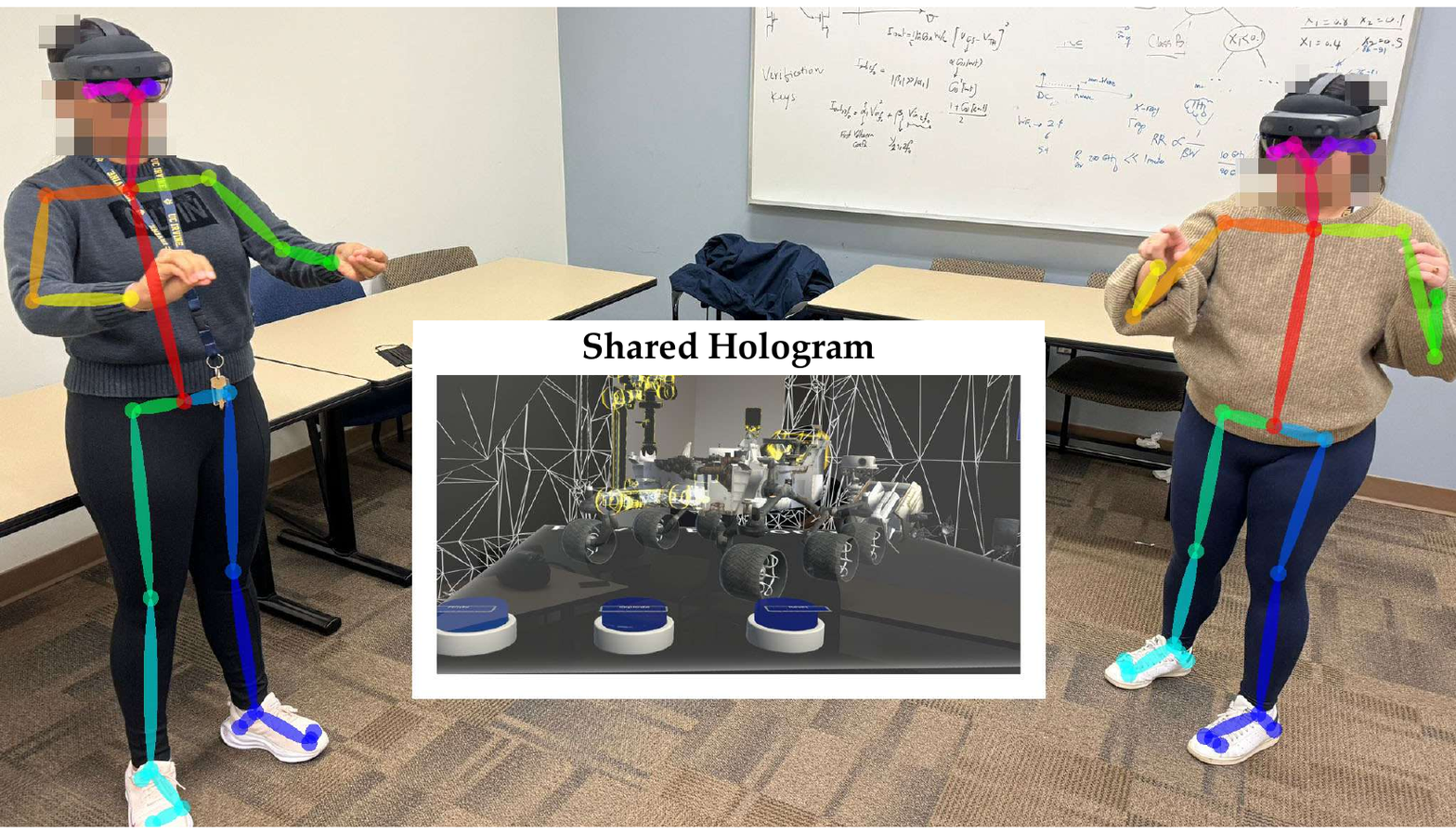} 
  \caption{Data collection setup for recording walking sequences using the HoloLens~2 within a collaborative MR  application. }  
  \label{fig:data_collection}
\end{figure}

\textbf{Data Collection.}
We conducted an IRB-approved user study with 20 participants (aged 18+), recruited via mailing lists and referrals. To protect identity, participants wore face masks during recording, and all data was anonymized. Participants walked between two markers (2.5--2.75m apart) for approximately 5 minutes while being recorded using the HoloLens~2 camera (Figure~\ref{fig:data_collection}). This setup reflects natural MR collaboration settings involving limited user movement. Our participant count aligns with prior AR privacy research (e.g., 16 participants in BystandAR~\cite{corbettBystandARProtectingBystander2023}); the goal was to evaluate the feasibility of gait profiling attacks and mitigation in typical MR use cases.

\textbf{Privacy Metrics.}
We applied the 233 configurations from Section~\ref{sec:mitigation} and used the fully automated \sysextract to extract gait features from original and mitigated video frames. This simulates a realistic adversarial scenario where the attacker has limited resources and no manual calibration access. We measure privacy using:

(1) \emph{Jensen-Shannon Divergence (JSD)} quantifies the statistical difference between original and mitigated feature distributions, ranging from 0 to 1. Higher values indicate greater feature distortion.

(2) \emph{Mann-Whitney U Test} determines whether original and mitigated features come from the same distribution. A p-value $< 0.05$ indicates statistically significant differences, suggesting effective mitigation.

(3) \emph{Gait Profiling Accuracy Reduction} measures the relative decrease in classifier performance after mitigation, computed as $(A_{\text{baseline}} - A_{\text{mitigated}}) /
  A_{\text{baseline}} \times 100\%$, where $A_{\text{baseline}}$ is the unmitigated profiling accuracy (more explanation in \S\ref{sec:baseline-attack}). For our 20-participant dataset, random-chance
  accuracy is 5\%; higher reduction values indicate the mitigated accuracy approaches this random baseline.

\textbf{Utility Metrics.}
To assess visual quality after mitigation, we used (1) \emph{Peak Signal-to-Noise Ratio (PSNR)} to evaluate distortion introduced across RGB channels and (2) \emph{Structural Similarity Index (SSIM)} to quantify perceived visual quality by comparing luminance, contrast, and structure between original and mitigated frames.

\textbf{System Performance Metrics.}\label{sec:system-perf}
On HoloLens~2, optimal application experience requires 60~fps~\cite{keveleighPerformanceMRTK2022}, while MR Capture typically achieves 30~fps for streaming~\cite{vtietoMixedRealityCapture2022}. We measured how mitigation processing affects application and streaming frame rates under static and adaptive configurations.

\subsection{Baseline Attack Performance}\label{sec:baseline-attack}

To establish the privacy threat, we first evaluate gait profiling accuracy without any mitigation. Using the \sysextract pipeline (\S\ref{sec:gaitextract}) to extract 10-dimensional gait features from video, a Gradient Boosting classifier achieves \textbf{78.2\% user profiling accuracy} across 20 participants---15.6$\times$ better than random guessing (5\%). The fully automated configuration using only 8 temporal features (excluding step length, which requires manual calibration) still achieves 69.9\% accuracy. Per-user analysis reveals substantial variance: accuracy ranges from 48.8\% to 92.3\% (see confusion matrix in Appendix~\S\ref{app:confusion-matrix}), indicating some users exhibit more distinctive gait patterns. Both configurations significantly outperform random chance ($p < 0.001$, binomial test). These results confirm that video-based gait analysis poses a significant privacy threat, motivating the mitigations evaluated below.

\begin{table}[t]
\footnotesize
\centering
\caption{Baseline Gait Profiling Attack Performance. Accuracy for association of gait features across 20 participants using Gradient Boosting with 6-fold cross-validation. Both configurations significantly outperform random baseline (5\%) with $p < 0.001$.}
\label{tab:baseline_metrics}
\begin{tabular}{lcc}
\toprule
\textbf{Metric} & \textbf{Without Step Length} & \textbf{With Step Length} \\
\midrule
Mean Accuracy & 69.9\% $\pm$ 0.4\% & 78.2\% $\pm$ 1.4\% \\
Mean F1 Score & 69.9\% $\pm$ 0.6\% & 78.1\% $\pm$ 1.4\% \\
Per-User Range & 35.4\%--90.6\% & 48.8\%--92.3\% \\
\bottomrule
\end{tabular}

\end{table}

\subsection{Mitigation Aspects Ablation Study}

We evaluated 233 unique mitigation configurations across 9 techniques and 5 target locations. Detailed results including per-configuration metrics (\S\ref{app:parameter-sweep}), distribution visualizations (\S\ref{app:distributions}), and the Pareto frontier analysis (\S\ref{app:pu-scatter}) are provided in Appendix~\ref{app:ablation}.

\begin{table}[t]
\centering
\caption{Metrics for Different Mitigation Locations.}
\label{tab:mitiloc_summary}
\setlength{\tabcolsep}{1pt}  
\footnotesize
    \begin{tabular}{lcccc}
    \toprule
    \textbf{\makecell[c]{Mitigation Location}} &\textbf{ \makecell[c]{Average \\ SSIM}} & \textbf{\makecell[c]{Average \\ PSNR}} & \textbf{\makecell[c]{Average \\ JSD}} & \textbf{\makecell[c]{Reduction in \\ Prof. Accuracy (\%)}} \\ \hline
    
    KPM & 0.957 & 32.060 & 0.408 & 8.9 \\
    
    \makecell[l]{Head and Shoulders KPM} & 0.958 & 32.850 & 0.456 & 6.9 \\ 
    
    \makecell[l]{Lower body KPM} & 0.955 & 32.349 & 0.495 & 19.9 \\ 
    
    \makecell[l]{All except head KPM} & 0.942 & 31.418 & 0.504 & 22.2 \\
    
    LBM & 0.925 & 24.835 & 0.580 & 56.6 \\
    \hline
    
    \end{tabular} 
    
\end{table}

\emph{Mitigation Location~(\ding{182}).}
As shown in Table~\ref{tab:mitiloc_summary}, baseline KPM and head \& shoulders KPM achieve high utility (SSIM$\approx$0.96, PSNR$\approx$32~dB) but minimal privacy ($<$10\% accuracy reduction, JSD$<$0.46). Targeting only a few keypoints inadequately protects gait privacy, possibly due to OpenPose's ability to interpolate obscured keypoints via part affinity fields~\cite{caoOpenPoseRealtimeMultiPerson2019}. Lower body KPM and all-except-head KPM improve privacy ($\approx$20\% accuracy reduction, JSD$\approx$0.50). LBM delivers the strongest privacy (56.6\% accuracy reduction, JSD: 0.58) but lowest utility (SSIM: 0.925, PSNR: 24.8~dB). These results emphasize the importance of targeting lower body regions; relying solely on standard keypoints is insufficient, and expanding beyond the lower body yields minimal additional benefit.

\begin{table}[t]
  \caption{Metrics for Different Masking Box Sizes}
\label{tab:mitisize}
\setlength{\tabcolsep}{1pt}  
\centering
\footnotesize
    \begin{tabular}{lcccc}
    \toprule
    \textbf{\makecell[c]{Box \\Size (px)}} &\textbf{ \makecell[c]{Average \\ SSIM}} & \textbf{\makecell[c]{Average \\ PSNR}} & \textbf{\makecell[c]{Average \\ JSD}} & \textbf{\makecell[c]{Reduction in \\ Prof. Accuracy (\%)}} \\ \hline
    
    50  & 0.979 & 35.193 &  5.167 & 0.457 \\
    100 & 0.955 & 31.430 & 21.889 & 0.478 \\
    150 & 0.926 & 29.163 & 24.032 & 0.489 \\
    200 & 0.926 & 28.037 & 20.634 & 0.457 \\
    \hline
  \end{tabular} 
  
\end{table}

\emph{Mitigation Size~(\ding{183}).}
We tested box sizes from 50 to 200 pixels, ranging from minimal point coverage to full joint occlusion (Table~\ref{tab:mitisize}). Smaller boxes (50~px) preserve quality (SSIM: 0.979, PSNR: 35.2~dB) but offer minimal privacy (5.2\% accuracy reduction). The 150-pixel configuration achieves peak privacy (JSD: 0.489, 24\% accuracy reduction) while maintaining acceptable quality (SSIM: 0.926). Increasing to 200~pixels does not further improve privacy, suggesting diminishing returns beyond a threshold, possibly due to overlap with non-informative regions. A 150-pixel box offers the most favorable trade-off.

\begin{table}[t]
\caption{Metrics for Different Mitigation Types}
\label{tab:mititype_summary}
\setlength{\tabcolsep}{1pt}  
\centering
\footnotesize
    \begin{tabular}{lcccc}
    \toprule
    \textbf{Type} & \textbf{\makecell[c]{Average \\SSIM}} & \textbf{\makecell[c]{Average \\PSNR}} & \textbf{\makecell[c]{Average \\JSD}} & \textbf{\makecell[c]{Reduction in \\Prof. Accuracy (\%)}} \\ \hline
    
    Black box & 0.930 & 23.570 & 0.516 & 31.915 \\
    
    DP Pix & 0.942 & 19.804 & 0.541 & 39.966 \\
    
    DP Laplace & 0.951 & 32.843 & 0.468 & 15.373 \\
    
    Blur & 0.981 & 37.865 & 0.380 & 4.492 \\
    
    Uniform & 0.958 & 34.167 & 0.441 & 20.103 \\
    
    Normal & 0.951 & 31.687 & 0.461 & 25.852 \\
    
    Laplace & 0.946 & 30.577 & 0.469 & 27.830 \\
    
    Exponential & 0.950 & 28.084 & 0.466 & 25.474 \\
    
    Average Color & 0.961 & 29.782 & 0.503 & 32.626 \\
    \hline
    
    \end{tabular}

\end{table}

\emph{Mitigation Type~(\ding{184}).}
We compare nine mitigation techniques in Table~\ref{tab:mititype_summary}. Black box offers strong privacy (31.9\% accuracy reduction, JSD: 0.516) but significant distortion (SSIM: 0.93, PSNR: 23.6~dB), suggesting the need for more nuanced alternatives. Random noise methods (uniform, normal, Laplace, exponential) provide moderate privacy (JSD: 0.44--0.47, 20--28\% accuracy reduction) with better utility. Despite its popularity, Gaussian blur shows weak privacy (JSD: 0.380, 4.5\% accuracy reduction), indicating ineffectiveness in concealing gait characteristics.

Two methods stand out: DP Pix achieves highest JSD (0.541) and accuracy reduction (40\%) but lower utility (PSNR: 19.8~dB). Average Color offers better balance: 32.6\% accuracy reduction, JSD of 0.503, with higher utility (SSIM: 0.961, PSNR: 29.8~dB). Among all methods, only black box, DP Pix, and Average Color produce significantly altered distributions (JSD~$>$0.5). Of these, Average Color offers the most compelling privacy-utility trade-off for real-time MR deployment.

Notably, applying DP pixelization or average color to lower body KPM with a 150-pixel box yields privacy and utility outcomes nearly identical to full LBM. This suggests that with sufficiently large masked regions, keypoint-based mitigation can approximate the effectiveness of full lower-body masking while offering greater modularity.

\begin{table}[t]
\caption{Comparison of Static and Adaptive Mitigation.}
\label{tab:mititemporal_summary}
\setlength{\tabcolsep}{1.5pt}
\centering
\footnotesize
\begin{tabular}{llcccc}
\toprule
\textbf{Type} & \textbf{Temporal} & \textbf{\makecell[c]{Average\\SSIM}} & \textbf{\makecell[c]{Average\\PSNR}} & \textbf{\makecell[c]{Reduction in\\Prof. Acc.}} & \textbf{\makecell[c]{Average\\JSD}} \\
\midrule
  Black Box & Static & 0.89 & 20.93 & 68\% & 0.63 \\
  Black Box & Adaptive & 0.95 & 23.50 & 68\% & 0.63 \\
  \midrule
  Ave.~Color & Static & 0.95 & 23.24 & 68\% & 0.63 \\
  Ave.~Color & Adaptive & 0.97 & 28.86 & 68\% & 0.63 \\
  \bottomrule
\end{tabular}

\end{table}

\subsection{Adaptive Mitigation Results}
  
To assess temporal adaptation, we compared static mitigation (applied to every frame) against adaptive mitigation (applied only to gait event frames). We selected the LBM location paired with two mitigation types: baseline black box and top-performing average color.

Privacy metrics remain consistent across all configurations: each achieves 68\% reduction in gait profiling accuracy and JSD of 0.63 (Table~\ref{tab:mititemporal_summary}). However, adaptive mitigation yields significant utility improvements. For average color, PSNR increases from 23.24~dB to 28.86~dB, a gain of over 5~dB, which represents a substantial improvement in visual quality. SSIM also improves from 0.95 to 0.97. Black box shows similar gains, with SSIM improving from 0.89 to 0.95 and PSNR from 20.93~dB to 23.50~dB.

These improvements stem from selective application of mitigation to gait-relevant frames (around heel-strike and toe-off events), reducing unnecessary distortion in \\non-informative frames. Adaptive mitigation maintains full privacy efficacy while significantly improving visual quality, making it especially valuable for real-time MR applications where user experience is critical.

\begin{table*}[!ht]
    \centering
    \caption{The privacy-utility tradeoff (PUT) of all mitigation techniques using gait features from the fully automated \sysextract. The privacy metric (reduction in gait profiling accuracy) was compared with two different utility metrics: (1) PSNR and (2) SSIM.}\label{tbl:combined-put-nsl-zero}
    
    \begin{tabularx}{\linewidth}{|c|Y|Y|Y|Y|Y|}
        \hline
        
        \multicolumn{6}{|c|}{\includegraphics[width=0.98\textwidth]{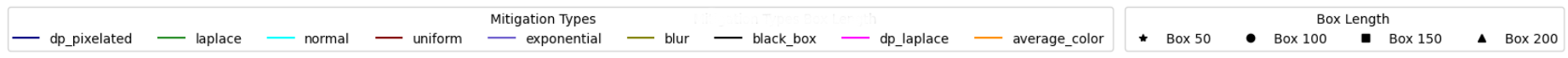}}  \\ \hline

        & 
        \textbf{Lower body-based Masking (LBM) } & \textbf{KPM} & \textbf{Lower Body KPM} & \textbf{Head and Shoulders KPM} & \textbf{All except head KPM} \\ \hline

        \rotatebox{90}{\textbf{Privacy vs.\ PSNR}} 
        & \includegraphics[width=0.18\textwidth]{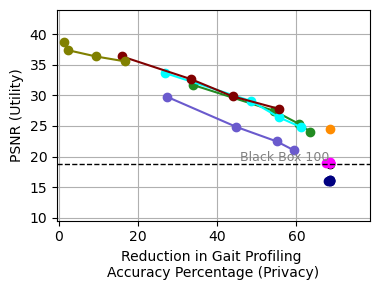} 
        & \includegraphics[width=0.18\textwidth]{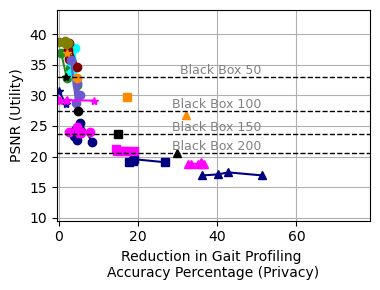} 
        & \includegraphics[width=0.18\textwidth]{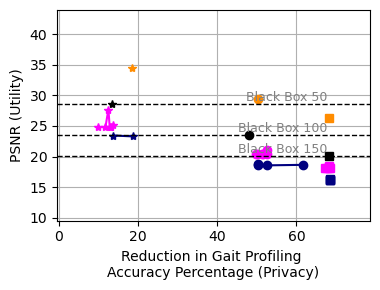} 
        & \includegraphics[width=0.18\textwidth]{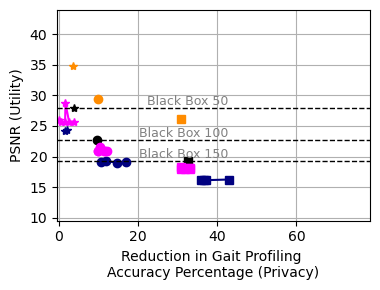} & \includegraphics[width=0.18\textwidth]{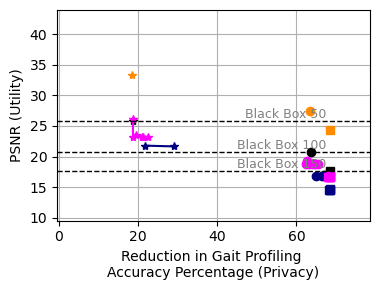} \\ \hline

        \rotatebox{90}{\textbf{Privacy vs.\ SSIM}} 
        & \includegraphics[width=0.18\textwidth]{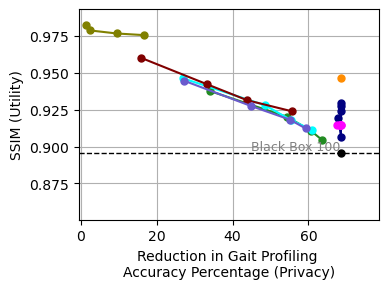} 
        & \includegraphics[width=0.18\textwidth]{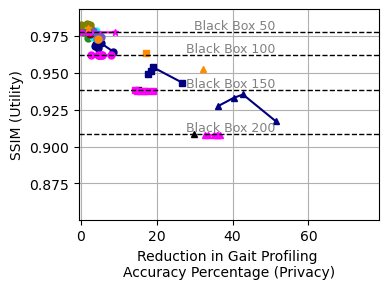} 
        & \includegraphics[width=0.18\textwidth]{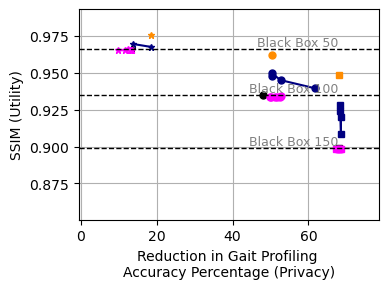} 
        & \includegraphics[width=0.18\textwidth]{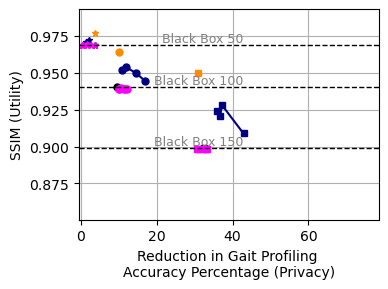} 
        & \includegraphics[width=0.18\textwidth]{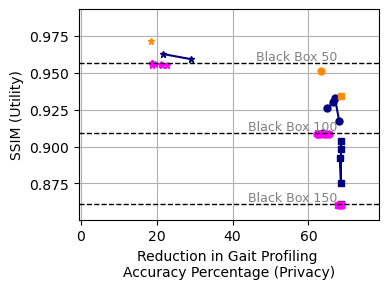} \\ \hline
    \end{tabularx} 
    
\end{table*}

\subsection{Temporal Attack Resilience}\label{sec:temporal-attack}

A potential concern with adaptive mitigation is that the \emph{timing} of mitigated frames could leak gait information. Since adaptive mitigation applies transformations only during gait events (heel-strike and toe-off), an adversary observing the output stream could identify mitigated frames and extract inter-event timing intervals for gait profiling.

\begin{figure}[t]
\centering
\includegraphics[width=\columnwidth]{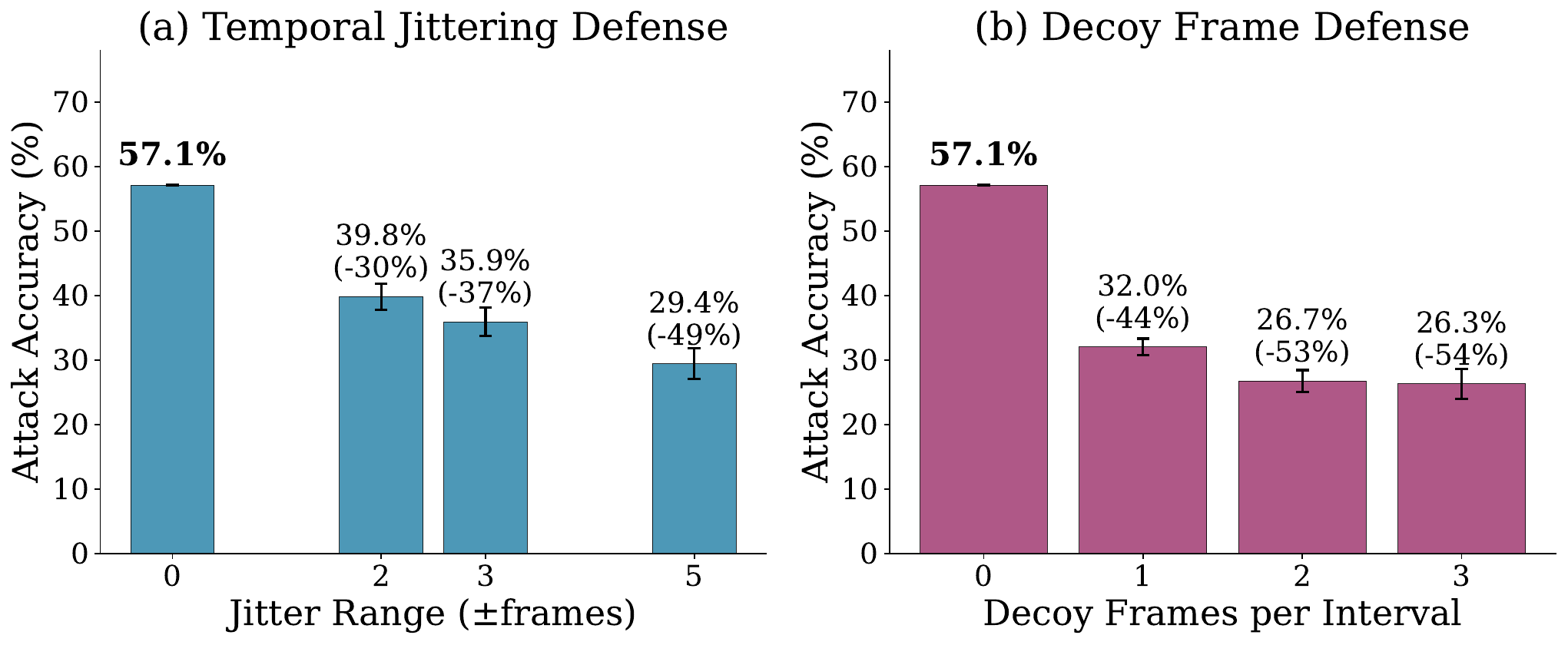}
\caption{Temporal side-channel analysis. (a)~Temporal jittering reduces profiling accuracy by 49\% at $\pm$5 frames. (b)~Decoy frames reduce accuracy by 53\% with 2 decoys per interval.}
\label{fig:temporal-attack}

\end{figure}

To evaluate this potential vulnerability, we trained a classifier on temporal features (inter-event intervals, step durations, cadence variability) extracted from 1,330 walking sequences across 20 users. As shown in Figure~\ref{fig:temporal-attack}, the baseline temporal attack achieves only 57.1\% profiling accuracy---substantially lower than visual gait recognition methods that achieve 80--90\% accuracy on our dataset. While above random chance (4.3\% for 20 users), temporal patterns alone provide limited profiling capability compared to visual analysis.

We evaluated two countermeasures to further reduce this residual vulnerability. \emph{Temporal jittering} adds random offsets ($\pm k$ frames) to mitigation timing, disrupting the correlation between mitigated frames and actual gait events. As shown in Figure~\ref{fig:temporal-attack}(a), $\pm$5 frames reduces profiling accuracy to 29.4\% (49\% reduction from baseline). \emph{Decoy frames} apply mitigation to additional non-gait-event frames, masking the true event timing. Figure~\ref{fig:temporal-attack}(b) shows that 2 decoy frames per interval reduce accuracy to 26.7\% (53\% reduction). Both countermeasures bring profiling accuracy to near-random levels with minimal computational overhead.

\subsection{Summary of Privacy-Utility Trade-offs}

We evaluated privacy-utility trade-offs across all 233 configurations by comparing gait profiling accuracy reduction against PSNR and SSIM. The results, visualized in Table~\ref{tbl:combined-put-nsl-zero}, reveal key findings. Average color mitigation at lower body KPM achieves the best overall performance, producing highest PSNR and SSIM while maintaining maximum observed privacy (68\% reduction in gait profiling accuracy). The same technique at LBM location achieves equivalent privacy but slightly lower utility (1.8~dB lower PSNR, 0.003 decrease in SSIM). Average color consistently offers the best privacy-utility trade-off across all locations, whereas black box, despite being equally effective for privacy, produces significantly worse utility. Baseline KPM targeting only three gait-related keypoints proves insufficient. OpenPose compensates for missing keypoints by inferring positions from surrounding body parts, reducing privacy impact. The head \& shoulders KPM configuration performs better by disrupting OpenPose's estimation more effectively. Applying DP pixelization or average color to LBM or lower body KPM with 150-pixel boxes yields similar outcomes. This confirms that KPM with sufficiently large masked regions can approximate full lower-body masking while offering greater modularity. Finally, adaptive strategies match static configurations for privacy while significantly improving utility. By focusing on gait-critical frames, adaptive methods reduce unnecessary distortion, making them ideal for real-time MR deployment.

\subsection{Information-Theoretic Analysis}

To provide theoretical grounding for our empirical results, we analyze information leakage from gait features to user identity using Mutual Information (MI). MI quantifies how many nats of information about identity can be extracted from gait data, independent of any specific classifier.

We computed MI between each gait feature and user identity across our 20-participant dataset (Figure~\ref{fig:mi_ranking} in Appendix~\S\ref{app:feature-mi}). Feature importance analysis
  using Gradient Boosting (Appendix~\S\ref{app:feature-imp}) corroborates these findings. The baseline (unmitigated) features contain 11.85~nats of total information about identity. Temporal features,  particularly step time and stance time, are most identifying (1.2--1.3~nats each), while spatial features (step length) leak substantially less information (0.48--0.49~nats). This aligns with our finding that lower-body masking, which obscures temporal gait dynamics, provides stronger privacy than keypoint-only approaches. Analyzing mitigation effectiveness through MI reveals that black box and DP pixelation achieve the highest information reduction (73--100\%) while blur provides only 19--27\% reduction, explaining why blur performs poorly in our empirical evaluation by failing to destroy the fine-grained temporal information that distinguishes individual gait patterns.

\subsection{Robustness Across Attack Methods}

To evaluate the generalization of \sysguard beyond our primary evaluation pipeline, we tested the best-performing configuration (average color at LBM) against two attacker methods representing different gait analysis approaches. The first attacker method uses \sysextract, our automated gait feature extraction pipeline (\S~\ref{sec:gaitextract}) that extracts clinical gait metrics from pose estimation. The second attacker method uses GaitLab~\cite{kidzinskiDeepNeuralNetworks2020}, a deep neural network that learns gait representations directly from video. These methods represent complementary attack strategies: pose-based clinical analysis versus end-to-end deep learning.

As shown in Table~\ref{table:gaitlab-comparison}, \sysguard significantly reduces gait profiling accuracy against both attacker methods: 64\% reduction for \sysextract and 36\% for GaitLab. JSD scores reflect the degree of feature distribution alteration: \sysextract shows substantial deviation (JSD: 0.63), while GaitLab shows moderate change (JSD: 0.22), consistent with its learned representations being partially robust to visual perturbations. Critically, Mann-Whitney U tests confirm statistically significant differences for both methods. These results demonstrate that \sysguard effectively disrupts both traditional and neural network-based gait analysis. The defense generalizes across attacker capabilities, establishing \sysguard as a practical solution for gait privacy protection regardless of the adversary's chosen analysis method.

\begin{table}[!t]
\caption{Cross-Method Robustness Evaluation}
\label{table:gaitlab-comparison}
\setlength{\tabcolsep}{4pt}  
\centering
\footnotesize
\begin{tabular}{lcc}\toprule
                               & \textbf{\sysextract} & {\textbf{Gait Lab}~\cite{kidzinskiDeepNeuralNetworks2020}} \\ 
\hline
Reduction in Prof. Accuracy                   & 64\%        & 36\%                                     \\

Average JSD                    & 0.63        & 0.22                                     \\

Significantly Different? (U-test) & Yes         & Yes  \\  
\hline
\end{tabular} 

\end{table}

\subsection{System Performance}

We evaluate \sysguard's real-time performance on a Samsung Galaxy S23 device, measuring frame rates, latency breakdown, and power consumption across static and adaptive configurations with varying ConvNet depths for gait event detection.

\begin{figure}[t]
  \centering
  \includegraphics[width=0.9\linewidth, trim=0 10 0 0, clip]{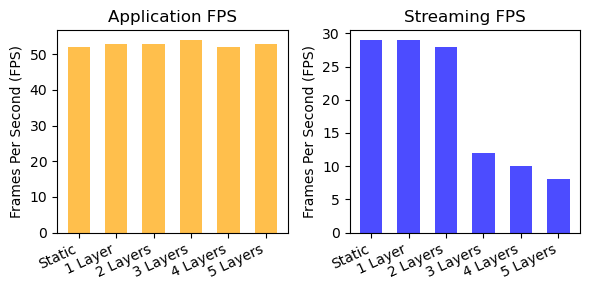}
  \caption{Performance comparison across static and adaptive configurations. Application FPS remains consistent (52--54), while streaming FPS decreases with model depth, showing the trade-off between complexity and efficiency.}
  \label{fig:performance-metrics}
  
\end{figure}

\textbf{Frame Rate Analysis.}
All configurations maintain high application FPS (52--54), indicating negligible overhead on MR responsiveness (Figure~\ref{fig:performance-metrics}). For streaming, static and single-layer adaptive configurations achieve 29~FPS, meeting real-time requirements. As model depth increases, streaming FPS drops significantly: 28~FPS with two layers, then sharply to 12, 10, and 8~FPS for three, four, and five layers respectively. Despite reduced frame rates, all adaptive configurations achieve over 90\% gait event recall and equivalent privacy protection (68\% accuracy reduction, JSD: 0.63). Since single-layer models match deeper variants in privacy efficacy while maintaining real-time streaming, one convolution layer suffices for typical indoor MR settings. Deeper models may benefit more challenging environments with occlusion, where improved gait event detection accuracy could offset the performance cost.

\begin{figure}[t]
\centering
\includegraphics[width=0.9\columnwidth]{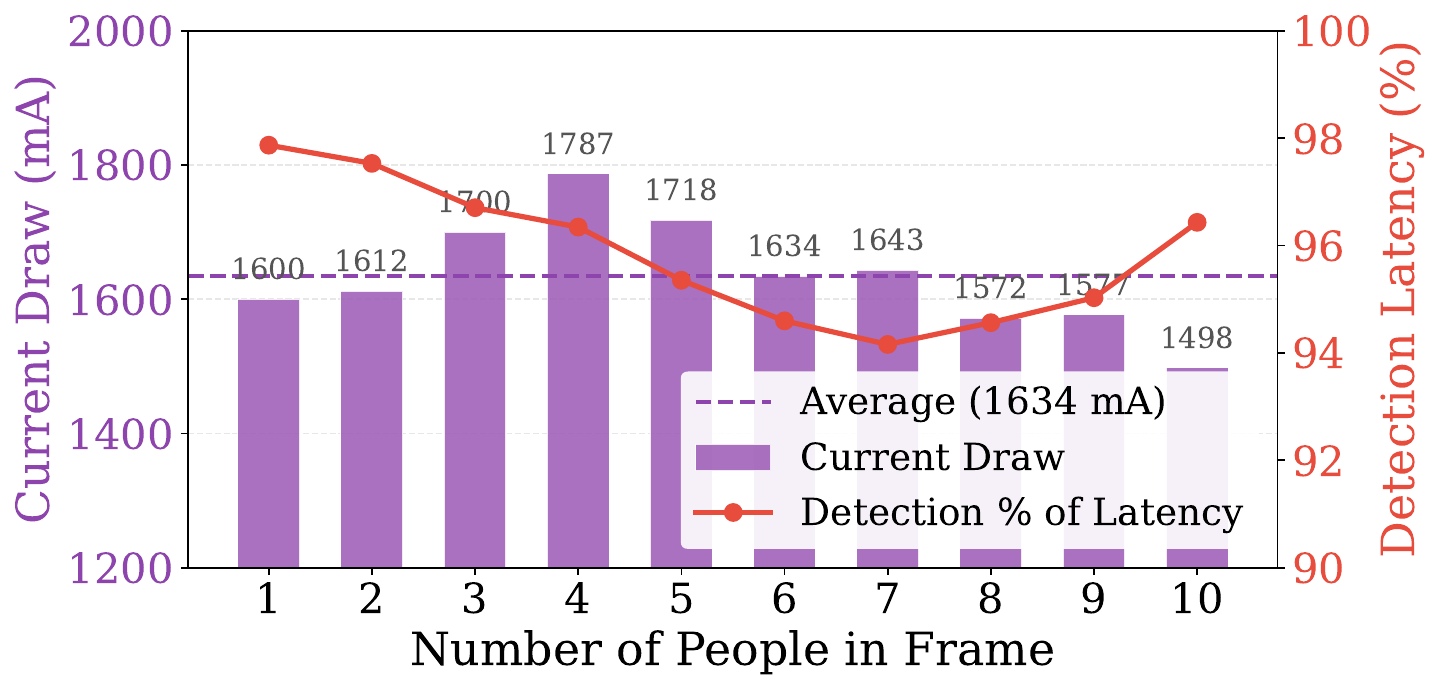}
\caption{Power consumption and per-module latency breakdown across person counts. Detection dominates both processing time (~96\%) and energy consumption, with mitigation overhead remaining negligible regardless of scene complexity.}
\label{fig:power}

\end{figure}

\textbf{Resource Consumption.}
We measured power consumption during video processing across 1--10 person configurations. Average current draw remained relatively constant at approximately 1.6~A regardless of person count, confirming that detection, not mitigation, dominates energy consumption (Figure~\ref{fig:power}). This aligns with our latency analysis showing detection accounts for 96\% of processing time. The consistent power profile demonstrates that \sysguard scales efficiently: adding more people to the scene incurs minimal additional energy cost.

\subsection{Scalability}

Most collaborative MR applications involve small groups. Enterprise use cases like remote assistance (Dynamics 365 Remote Assist~\cite{microsoft_dynamics_remote_assist_overview}), design review (Trimble Connect~\cite{trimble_connect_mr}), and creative work (Gravity Sketch~\cite{gravity_sketch}) typically support 1--5 users due to requirements for high-fidelity interaction, clear communication, and manageable cognitive load. Prior work on MR group behavior similarly focuses on 3--6 members~\cite{lowry2006impact,romero2024groupbeamr}. We evaluated \sysguard's per-frame processing latency with 1--10 people in the camera's field of view, measuring detection and mitigation time separately (see Figure~\ref{fig:scalability} in Appendix~\S\ref{app:multi-latency}). Detection time (HOG person detection) remains constant at approximately 8.4ms regardless of person count, as the detector scans the entire frame once. Mitigation time scales linearly with the number of detected people, adding roughly 0.05ms per person for the average-color transformation. Total processing latency ranges from 8.5ms (1 person) to 9ms (10 people), remaining well below the 33ms threshold required for 30~FPS real-time streaming~\cite{milazzo2021fps}. This provides approximately 25ms of headroom for network transmission and rendering overhead. The results confirm that detection---not mitigation---is the computational bottleneck, accounting for 96\% of total latency. Consequently, \sysguard scales efficiently: even in crowded scenes with 10 people, the system maintains real-time performance with minimal degradation.

\subsection{Qualitative Evaluation}

\begin{figure}[!t]
    \centering
    \includegraphics[width=1\linewidth]{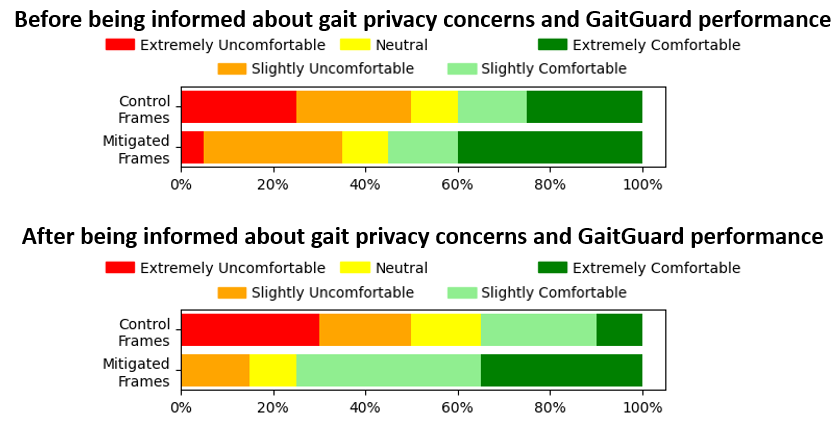}
    \caption{
    The Likert responses for the gait privacy perception survey were distributed across unmodified and mitigated frames with \sysguard. }
    \label{fig:survey}
\end{figure}

We conducted a two-part user study with 20 participants (balanced for VR/MR experience) to evaluate perceptions of gait privacy. Participants rated comfort in sharing video frames using a 5-point Likert scale (1 = extreme discomfort, 5 = extreme comfort). Initially, 50\% expressed discomfort sharing unmitigated video, while 40\% reported comfort. After viewing \sysguard-processed frames, comfort rose to 65\% (Figure~\ref{fig:survey}). In the second phase, participants were informed about gait privacy risks (inference of age, gender, health conditions) and \sysguard's performance (78\% to 16\% profiling accuracy, 0.123s delay). Risk awareness decreased comfort with raw footage, but 75\% reported comfort sharing \sysguard-processed video.

These findings indicate that both technological intervention and risk awareness shape privacy perceptions. Participants favored solutions providing strong privacy guarantees without degrading utility, supporting \sysguard as a practical, user-aligned approach to gait privacy in MR.

%% file: sections/6-discussion.tex
\section{Discussion}\label{sec:discussion}

\noindent\textbf{Broader Applicability.}
While we target MR environments where bystanders are exposed to always-on cameras, \sysguard applies to any system with video capture: smart doorbells, surveillance cameras, or video conferencing. The core insight that gait information concentrates in sparse temporal events generalizes across these domains. Our companion-based architecture can be adapted to edge processing nodes or cloud ingestion points depending on constraints.

\noindent\textbf{Defense Strategy.}

Unlike adversarial perturbations, which are classifier-specific and susceptible to retraining, \sysguard employs geometric obfuscation to remove discriminative gait information directly from the video stream. Evaluation against two distinct attack methods (pose-based GaitExtract and neural GaitLab) demonstrates effectiveness across analysis paradigms, though the degree of protection varies (64\% vs 36\% accuracy reduction), reflecting differences in feature representations.

\noindent\textbf{Privacy-Utility Tradeoffs.}
Our Pareto analysis reveals that average color masking at lower-body regions achieves near-optimal privacy with minimal utility loss. Applications can tune this tradeoff: security-critical scenarios may prefer\\ stronger protection (black box masking), while consumer applications may prioritize visual quality. The adaptive mitigation mode offers a third option, matching static privacy protection while improving utility by 5+ dB PSNR.

\noindent\textbf{Hardware Generalization.}
Our prototype targets HoloLens~2 with an S23 companion, but imposes minimal hardware assumptions. \sysguard processes standard RGB frames at 640$\times$480 resolution; any device exposing a camera API can serve as input. As MR hardware integrates dedicated neural accelerators, \sysguard can migrate to fully on-device processing. Our companion app reflects current platform constraints rather than fundamental design choices.

\noindent\textbf{Limitations and Future Work.}
Our evaluation used controlled indoor environments with 20 participants. Real deployments will have diverse lighting, occlusions, and crowds. However, \sysguard's robustness derives from YOLO and OpenPose, validated across challenging conditions, and our gait event CNN operating on pose skeletons rather than raw pixels. The temporal sparsity insight assumes gait events are reliably detectable; scenarios with severe occlusion or non-standard gaits (e.g., mobility aids) may degrade detection accuracy. Field studies remain important future work.

\noindent\textbf{User Perception.}
Our qualitative study revealed that informed consent significantly affects user comfort with privacy-preserving video. Technical mechanisms alone are insufficient; deployments should incorporate user education about threat models and the protections provided.

%% file: sections/7-conclusion.tex
\section{Conclusion}\label{sec:conclusion}

We presented \sysguard, a real-time system for mitigating gait privacy risks in MR environments. Central to our approach is \sysextract, a fully automated pipeline that extracts clinical-grade gait features from video without manual calibration, enabling both the quantification of profiling threats and the evaluation of defenses. Our evaluation demonstrates that gait-based profiling achieves 78\% accuracy on unprotected video (15.6$\times$ random chance), confirming the severity of this threat. Through systematic analysis of 233 mitigation configurations, we show that adaptive average color masking reduces profiling accuracy by 68\% while preserving visual quality (SSIM: 0.97). \sysguard~maintains 29~FPS streaming on commodity mobile hardware and scales to 10 simultaneous users with under 10ms total processing latency. By processing video on a companion device before transmission to external servers or applications, \sysguard~provides practical, immediately deployable protection against gait-based profiling in collaborative MR scenarios.

%% file: sections/8-appendix.tex
\savebox{\wideFigBox}{
  \begin{minipage}{\textwidth}
    \centering
    \includegraphics[width=\linewidth, trim=0 335 10 145, clip]{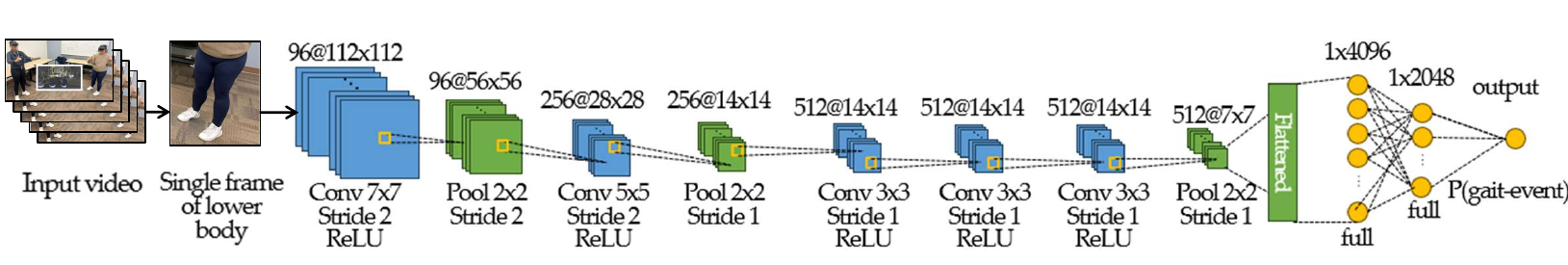}
    \captionof{figure}{Spatial ConvNet architecture for classifying gait events, providing the likelihood of a frame containing a heel strike or toe-off event.}
    \label{fig:gait-convnet}
  \end{minipage}
}
\twocolumn[\usebox{\wideFigBox}]

\section*{Appendix}

\section{Gait Event Classification Architecture}\label{app:convnet}

The spatial ConvNet classifies individual video frames as containing gait events (heel strike or toe-off) or non-events. The architecture, as visualized in Figure~\ref{fig:gait-convnet} operates on 18-joint pose skeletons extracted by OpenPose, represented as 36-dimensional vectors (x, y coordinates per joint).

The network consists of three convolutional blocks with increasing filter depths (64, 128, 256), each followed by batch normalization and ReLU activation. Max pooling reduces spatial dimensions between blocks. A global average pooling layer feeds into two fully connected layers (512 and 128 units) with dropout (p=0.5) for regularization. The final softmax layer outputs probabilities for three classes: heel strike, toe-off, and non-event.

Training used cross-entropy loss with class weights inversely proportional to frequency, addressing the imbalance where gait events constitute only 8\% of frames. The model achieves 94.2\% precision and 91.7\% recall on held-out test data, sufficient for reliable temporal sparsity exploitation in the mitigation pipeline.

\section{Detailed Mitigation Ablation Analysis}\label{app:ablation}

This appendix provides granular results from the ablation study summarized in Section~\ref{sec:evaluation}. We evaluated 233 unique mitigation configurations spanning 9 techniques, 5 target locations, and varying parameters.

\subsection{Privacy-Utility Tradeoff Visualization}\label{app:pu-scatter}

\begin{figure}[H]
\centering
\includegraphics[width=\columnwidth]{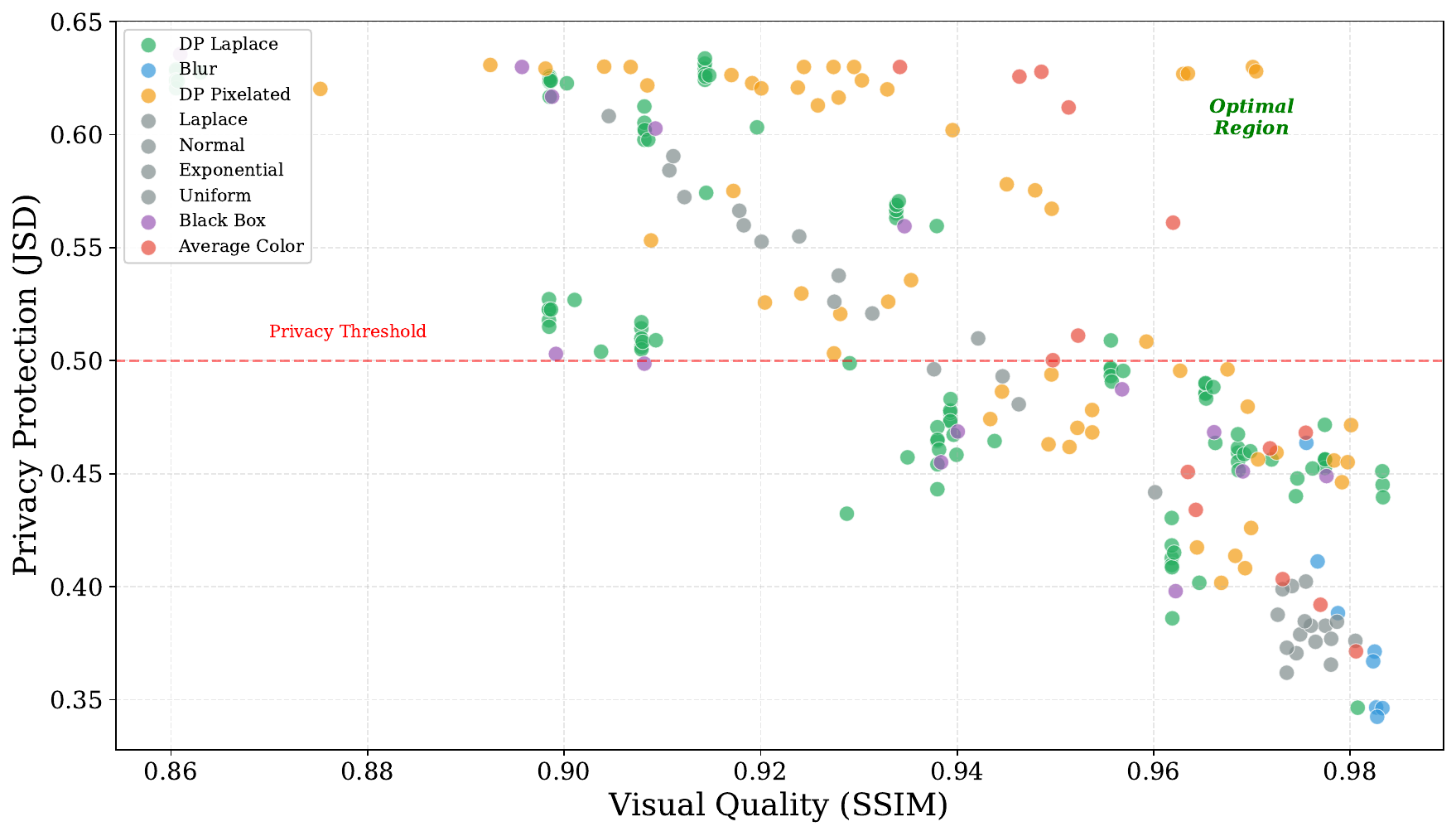}
\caption{Privacy (JSD) versus utility (SSIM) for all evaluated configurations. The horizontal dashed line at JSD=0.5 marks the threshold for meaningful privacy protection. Configurations in the upper-right region achieve optimal tradeoffs.}
\label{fig:pu_scatter}
\end{figure}

Figure~\ref{fig:pu_scatter} visualizes all configurations, revealing the Pareto frontier of privacy-utility tradeoffs. Each point represents a unique combination of mitigation type, location, and parameters.

Pareto-optimal configurations achieve SSIM $>$ 0.95 while maintaining JSD $>$ 0.5. DP Pixelated with lower lambda values and Average Color at lower-body locations occupy this region.

\subsection{Metric Distributions by Mitigation Type}\label{app:distributions}

\begin{figure}[H]
\centering
\includegraphics[width=\columnwidth]{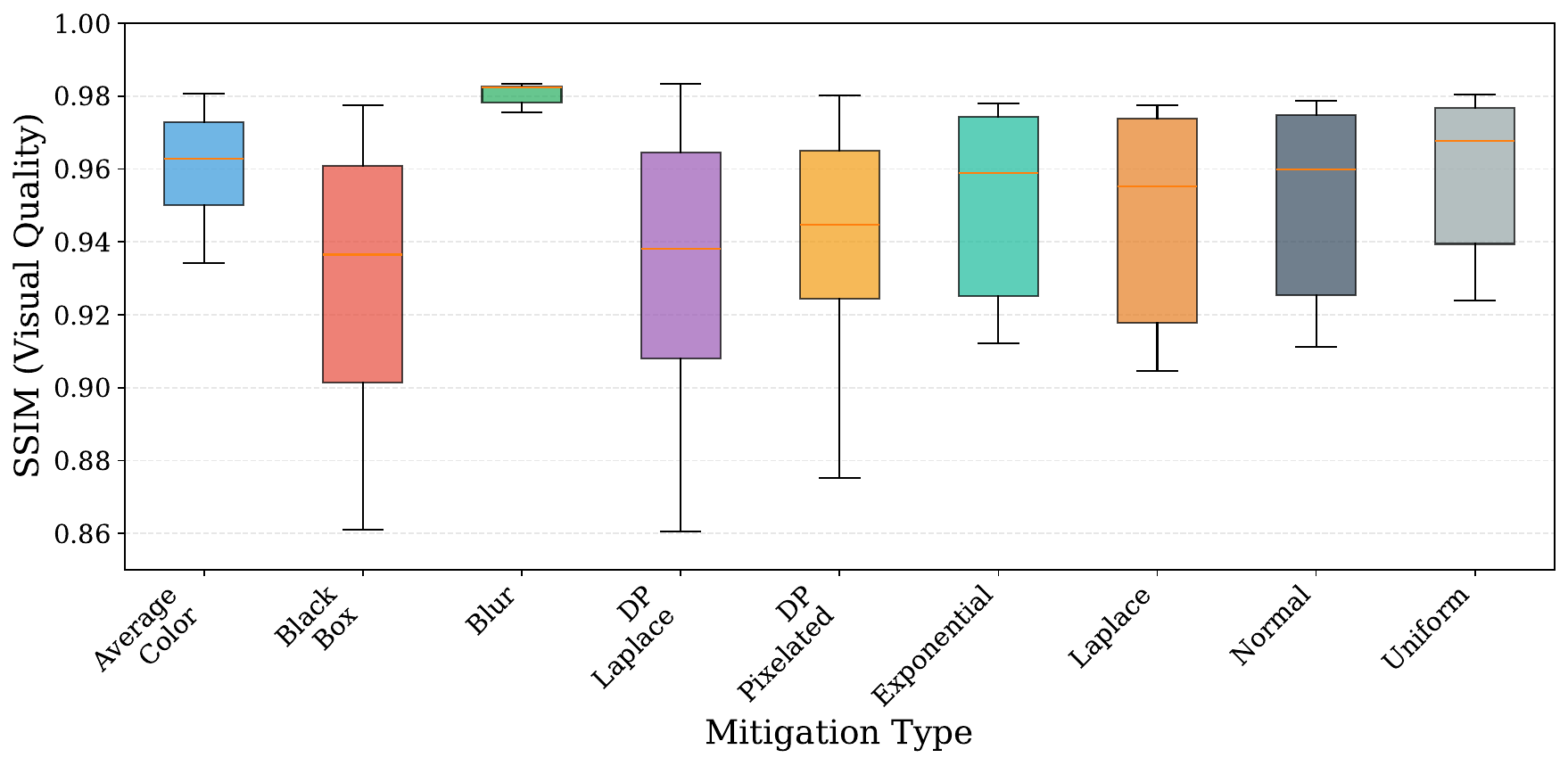}
\caption{SSIM distribution across configurations for each mitigation type. Blur methods achieve highest median SSIM with minimal variance. DP methods show greater variance depending on parameter settings.}
\label{fig:ssim_boxplots}
\end{figure}

\begin{figure}[H]
\centering
\includegraphics[width=\columnwidth]{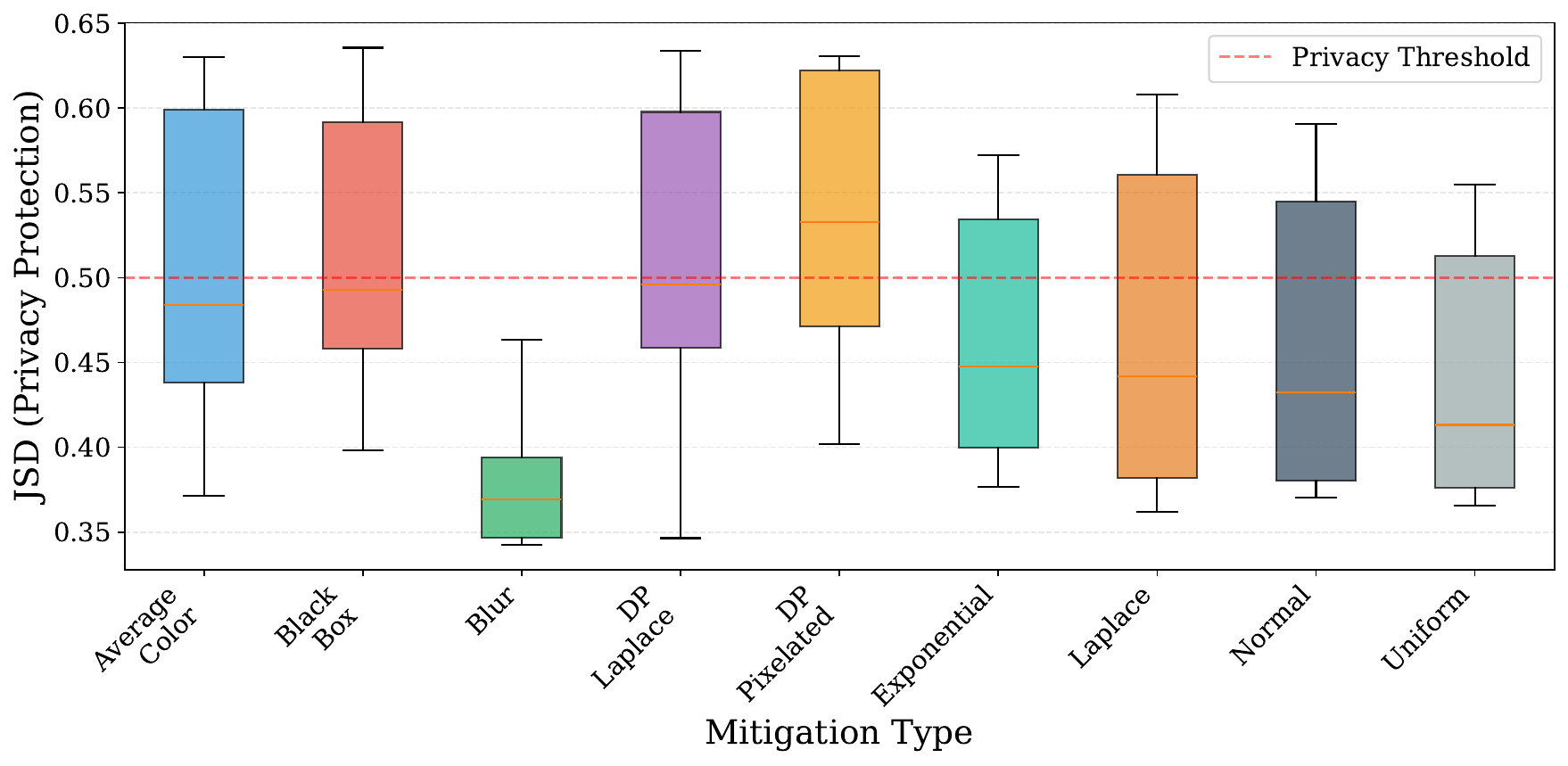}
\caption{JSD distribution across configurations for each mitigation type. The horizontal dashed line marks the JSD=0.5 privacy threshold. Average Color, Black Box, and DP Pixelated consistently achieve distributions above this threshold.}
\label{fig:jsd_boxplots}
\end{figure}

Visual quality and privacy protection vary across mitigation types. Figure~\ref{fig:ssim_boxplots} shows SSIM distributions, while Figure~\ref{fig:jsd_boxplots} shows JSD distributions for each mitigation category.

\subsection{Complete Parameter Sweep Results}\label{app:parameter-sweep}

Table~\ref{tab:parameter_sweep_full} presents the complete results for all 233 configurations evaluated in the ablation study.

\section{Per-Participant Analysis}

\subsection{Baseline Profiling Confusion Matrix}\label{app:confusion-matrix}

\begin{figure}[H]
\centering
\includegraphics[width=0.85\columnwidth]{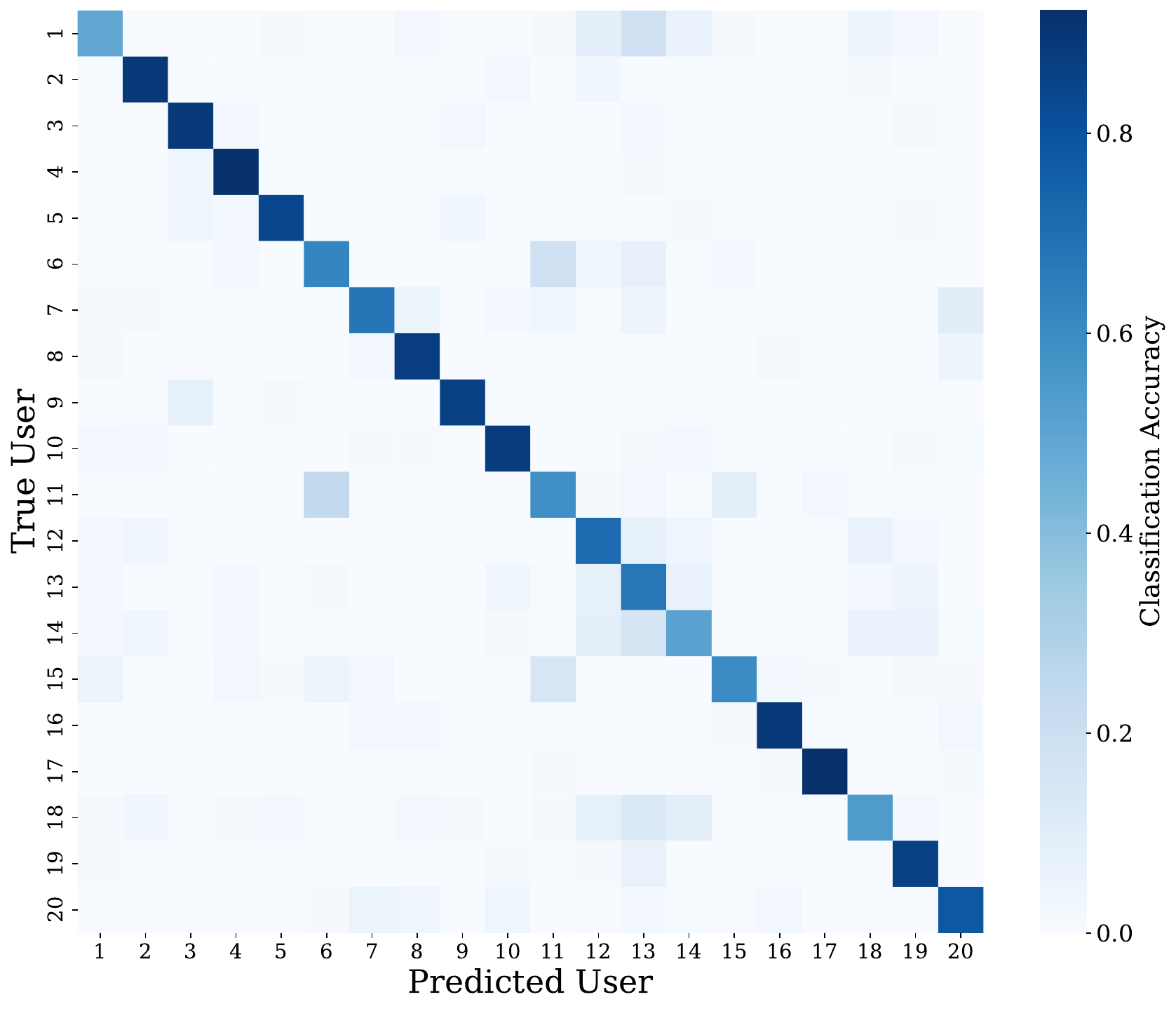}
\caption{20$\times$20 confusion matrix for baseline gait profiling. Strong diagonal indicates successful gait profiling. Off-diagonal values show misclassification patterns between similar gaits.}
\label{fig:confusion_matrix}
\end{figure}

The confusion matrix in Figure~\ref{fig:confusion_matrix} reveals which users are most distinguishable (strong diagonal) versus commonly confused pairs. Mean accuracy: 78.2\% $\pm$ 1.4\% across 20 participants.

\subsection{Gait Feature Importance}\label{app:feature-imp}

\begin{figure}[H]
\centering
\includegraphics[width=\columnwidth]{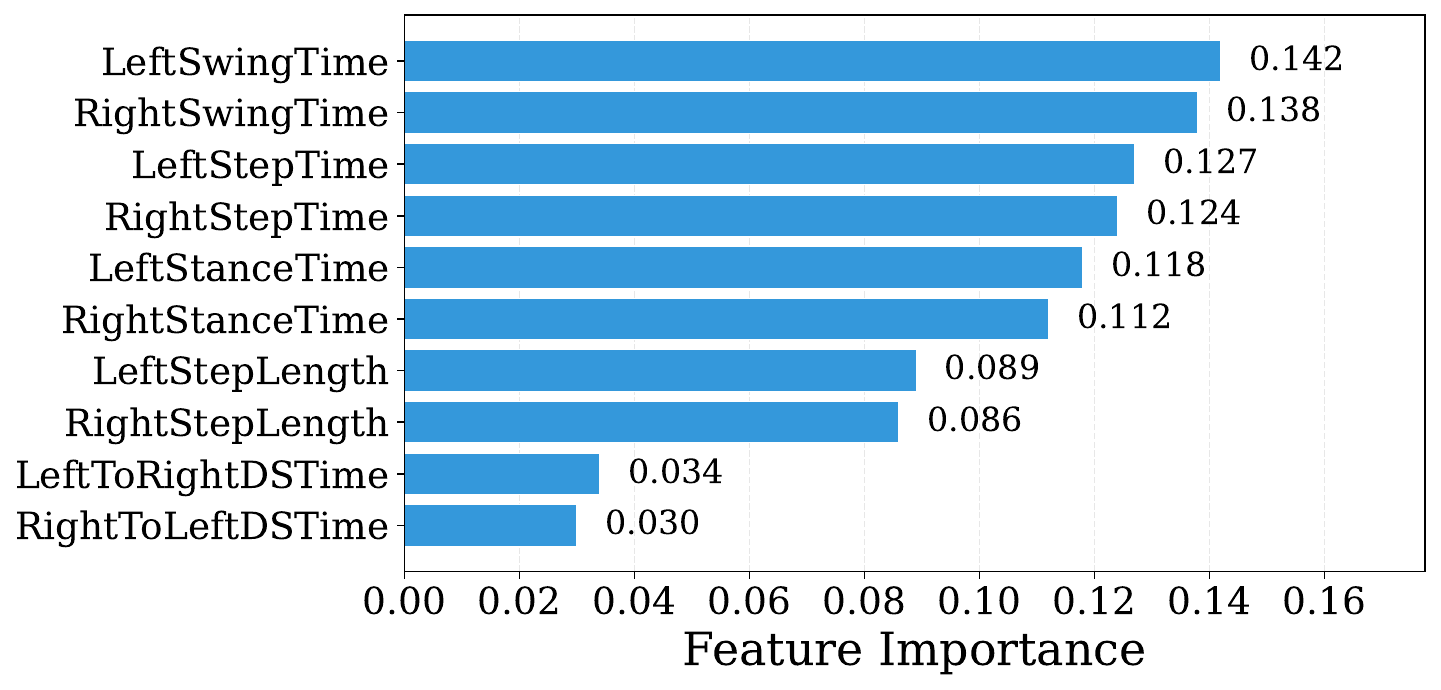}
\caption{Gait feature importance from Gradient Boosting classifier. Swing time and step length dominate, while double-support time contributes least.}
\label{fig:feature_importance}
\end{figure}

Feature importance analysis, summarized in Figure~\ref{fig:feature_importance} reveals which gait parameters contribute most to profiling, informing mitigation design.

\section{Per-Feature Mutual Information}\label{app:feature-mi}

\begin{figure}[H]
\centering
\includegraphics[width=\columnwidth]{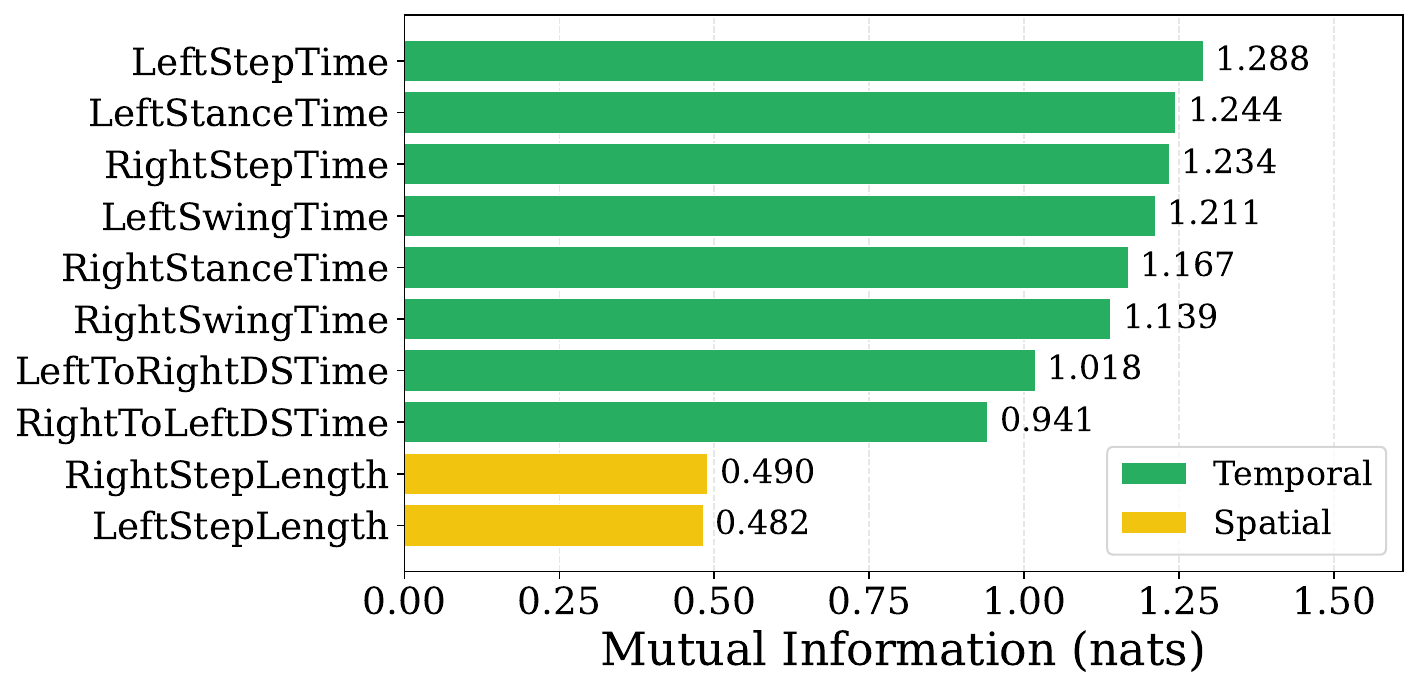}
\caption{Per-feature mutual information with user identity. Temporal features (step/stance/swing times) leak more information than spatial features (step length).}
\label{fig:mi_ranking}
\end{figure}

We quantify information leakage using mutual information (MI) between each gait feature and user identity and is shown in Figure~\ref{fig:mi_ranking}. Total baseline: 11.85 nats. 

\section{Multi-Person Latency Breakdown}\label{app:multi-latency}

\begin{figure}[H]
\centering
\includegraphics[width=0.85\columnwidth]{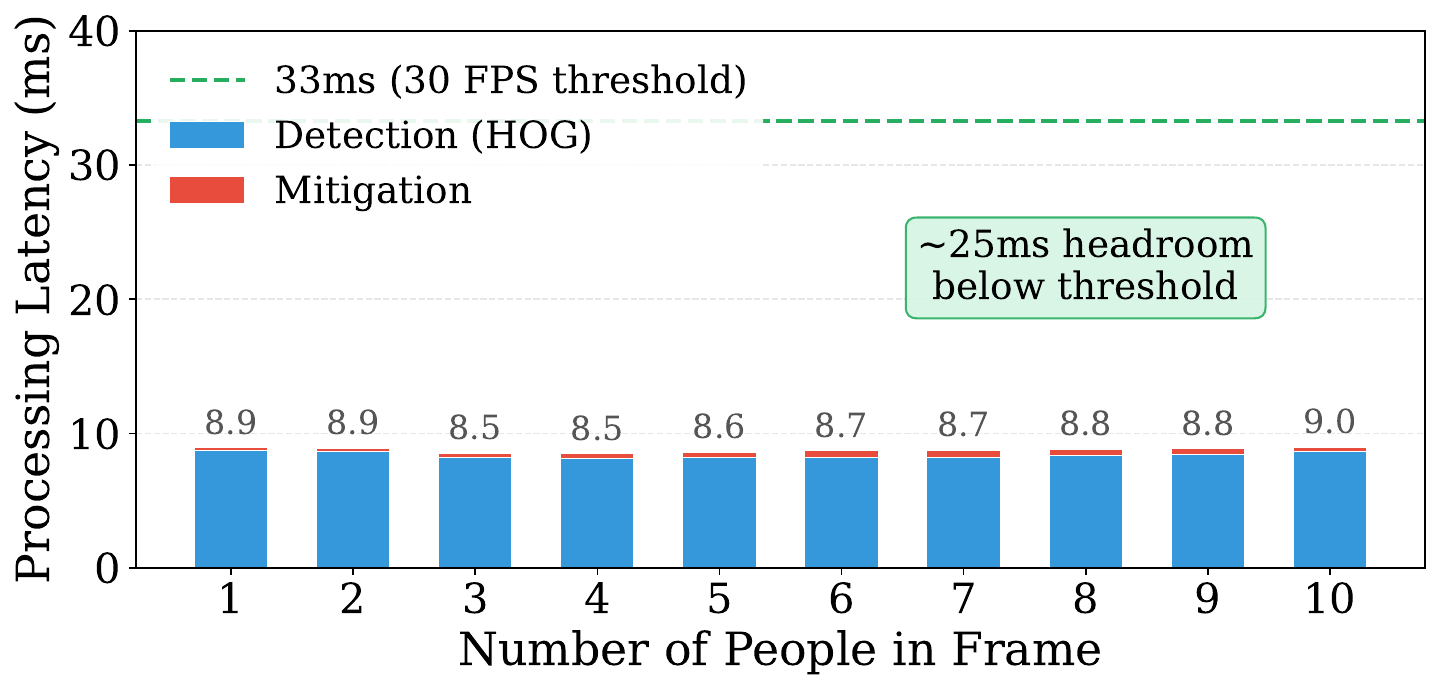}
\caption{Per-module latency with 1--10 people. Detection (96\% of processing) is constant; mitigation overhead is negligible. System maintains $>$100 FPS across all configurations.}
\label{fig:scalability}
\end{figure}

Detection time remains constant regardless of person count; mitigation scales linearly at $\sim$0.05 ms/person as seen in Figure~\ref{fig:scalability}.

 \onecolumn   
  \input{tables/parameter_sweep_table}
  \twocolumn   

%% file: tables/parameter_sweep_table.tex
\onecolumn
\small
\begin{longtable}{llrrcc|llrrcc}
\caption{Parameter Sweep Results (233 Configurations). Left and right columns show consecutive configurations sorted by mitigation type and JSD.}
\label{tab:parameter_sweep_full} \\
\hline
\multicolumn{6}{c|}{\textbf{Configurations 1--117}} & \multicolumn{6}{c}{\textbf{Configurations 118--233}} \\
\textbf{Type} & \textbf{Loc} & \textbf{Sz} & \textbf{$\lambda$} & \textbf{JSD} & \textbf{Acc} &
\textbf{Type} & \textbf{Loc} & \textbf{Sz} & \textbf{$\lambda$} & \textbf{JSD} & \textbf{Acc} \\
\hline
\endfirsthead
\multicolumn{12}{c}{\tablename\ \thetable{} -- continued} \\
\hline
\textbf{Type} & \textbf{Loc} & \textbf{Sz} & \textbf{$\lambda$} & \textbf{JSD} & \textbf{Acc} &
\textbf{Type} & \textbf{Loc} & \textbf{Sz} & \textbf{$\lambda$} & \textbf{JSD} & \textbf{Acc} \\
\hline
\endhead
\hline
\endfoot
\hline
\endlastfoot
Ave. Color & AEH-KPM & 150 & - & 0.63 & 69\% & DP Lap. & KPM & 150 & 100000.0 & 0.46 & 15\% \\
Ave. Color & LB-KPM & 150 & - & 0.63 & 68\% & DP Lap. & SH-KPM & 150 & 1000000.0 & 0.46 & 7\% \\
Ave. Color & LBM & 100 & - & 0.63 & 69\% & DP Lap. & KPM & 50 & 0.1 & 0.46 & 2\% \\
Ave. Color & AEH-KPM & 100 & - & 0.61 & 63\% & DP Lap. & KPM & 50 & 1000.0 & 0.46 & -0\% \\
Ave. Color & LB-KPM & 100 & - & 0.56 & 50\% & DP Lap. & KPM & 50 & 10.0 & 0.46 & 0\% \\
Ave. Color & KPM & 200 & - & 0.51 & 32\% & DP Lap. & LB-KPM & 100 & 1000000.0 & 0.46 & 4\% \\
Ave. Color & SH-KPM & 150 & - & 0.50 & 31\% & DP Lap. & SH-KPM & 50 & 100.0 & 0.46 & 4\% \\
Ave. Color & LB-KPM & 50 & - & 0.47 & 19\% & DP Lap. & KPM & 150 & 0.1 & 0.45 & 18\% \\
Ave. Color & AEH-KPM & 50 & - & 0.46 & 19\% & DP Lap. & KPM & 50 & 100.0 & 0.45 & 2\% \\
Ave. Color & KPM & 150 & - & 0.45 & 17\% & DP Lap. & SH-KPM & 50 & 100000.0 & 0.45 & 0\% \\
Ave. Color & SH-KPM & 100 & - & 0.43 & 10\% & DP Lap. & SH-KPM & 50 & 1000.0 & 0.45 & 0\% \\
Ave. Color & KPM & 100 & - & 0.40 & 5\% & DP Lap. & AEH-KPM & 50 & 1000000.0 & 0.45 & -0\% \\
Ave. Color & SH-KPM & 50 & - & 0.39 & 4\% & DP Lap. & LB-KPM & 50 & 100000.0 & 0.45 & 5\% \\
Ave. Color & KPM & 50 & - & 0.37 & 2\% & DP Lap. & LB-KPM & 50 & 1000000.0 & 0.45 & 1\% \\
 & & & & & & DP Lap. & KPM & 150 & 1.0 & 0.44 & 15\% \\
\cline{1-6}
Black Box & AEH-KPM & 150 & - & 0.64 & 69\% & DP Lap. & SH-KPM & 100 & 1000000.0 & 0.44 & 2\% \\
Black Box & LBM & 100 & - & 0.63 & 69\% & DP Lap. & SH-KPM & 50 & 1000000.0 & 0.44 & 1\% \\
Black Box & LB-KPM & 150 & - & 0.62 & 68\% & DP Lap. & KPM & 200 & 1000000.0 & 0.43 & 14\% \\
Black Box & AEH-KPM & 100 & - & 0.60 & 64\% & DP Lap. & KPM & 100 & 0.1 & 0.43 & 5\% \\
Black Box & LB-KPM & 100 & - & 0.56 & 48\% & DP Lap. & KPM & 100 & 10.0 & 0.42 & 6\% \\
Black Box & SH-KPM & 150 & - & 0.50 & 33\% & DP Lap. & KPM & 100 & 10000.0 & 0.42 & 5\% \\
Black Box & KPM & 200 & - & 0.50 & 30\% & DP Lap. & KPM & 100 & 1.0 & 0.41 & 6\% \\
Black Box & AEH-KPM & 50 & - & 0.49 & 19\% & DP Lap. & KPM & 100 & 100.0 & 0.41 & 3\% \\
Black Box & SH-KPM & 100 & - & 0.47 & 10\% & DP Lap. & KPM & 100 & 1000.0 & 0.41 & 8\% \\
Black Box & LB-KPM & 50 & - & 0.47 & 13\% & DP Lap. & KPM & 100 & 100000.0 & 0.40 & 6\% \\
Black Box & KPM & 150 & - & 0.46 & 15\% & DP Lap. & KPM & 150 & 1000000.0 & 0.39 & 4\% \\
Black Box & SH-KPM & 50 & - & 0.45 & 4\% & DP Lap. & KPM & 100 & 1000000.0 & 0.35 & 1\% \\
Black Box & KPM & 50 & - & 0.45 & 2\% & & & & & & \\
\cline{7-12}
Black Box & KPM & 100 & - & 0.40 & 5\% & DP Pix. & AEH-KPM & 150 & 20.0 & 0.63 & 68\% \\
 & & & & & & DP Pix. & AEH-KPM & 150 & 40.0 & 0.63 & 69\% \\
\cline{1-6}
Blur & LBM & 100 & 65.0 & 0.46 & 17\% & DP Pix. & LBM & 100 & 10.0 & 0.63 & 69\% \\
Blur & LBM & 100 & 45.0 & 0.41 & 9\% & DP Pix. & LBM & 100 & 30.0 & 0.63 & 69\% \\
Blur & LBM & 100 & 25.0 & 0.39 & 2\% & DP Pix. & LBM & 100 & 40.0 & 0.63 & 69\% \\
Blur & KPM & 100 & 45.0 & 0.37 & 2\% & DP Pix. & LBM & 100 & 50.0 & 0.63 & 69\% \\
Blur & KPM & 100 & 65.0 & 0.37 & 2\% & DP Pix. & LB-KPM & 50 & 30.0 & 0.63 & 69\% \\
Blur & LBM & 100 & 5.0 & 0.35 & 1\% & DP Pix. & AEH-KPM & 150 & 30.0 & 0.63 & 69\% \\
Blur & KPM & 100 & 5.0 & 0.35 & 2\% & DP Pix. & LB-KPM & 50 & 40.0 & 0.63 & 69\% \\
Blur & KPM & 100 & 25.0 & 0.34 & 0\% & DP Pix. & AEH-KPM & 50 & 40.0 & 0.63 & 69\% \\
 & & & & & & DP Pix. & AEH-KPM & 50 & 30.0 & 0.63 & 69\% \\
\cline{1-6}
DP Lap. & LBM & 100 & 1000.0 & 0.63 & 69\% & DP Pix. & AEH-KPM & 100 & 10.0 & 0.63 & 68\% \\
DP Lap. & LBM & 100 & 10.0 & 0.63 & 68\% & DP Pix. & AEH-KPM & 100 & 30.0 & 0.62 & 67\% \\
DP Lap. & LBM & 100 & 1.0 & 0.63 & 69\% & DP Pix. & LBM & 100 & 20.0 & 0.62 & 68\% \\
DP Lap. & AEH-KPM & 150 & 10.0 & 0.63 & 69\% & DP Pix. & LB-KPM & 150 & 10.0 & 0.62 & 69\% \\
DP Lap. & AEH-KPM & 150 & 0.1 & 0.63 & 69\% & DP Pix. & LB-KPM & 150 & 30.0 & 0.62 & 68\% \\
DP Lap. & AEH-KPM & 150 & 100000.0 & 0.63 & 68\% & DP Pix. & LB-KPM & 150 & 20.0 & 0.62 & 69\% \\
DP Lap. & LBM & 100 & 100.0 & 0.63 & 68\% & DP Pix. & AEH-KPM & 150 & 10.0 & 0.62 & 69\% \\
DP Lap. & LBM & 100 & 100000.0 & 0.63 & 68\% & DP Pix. & AEH-KPM & 100 & 40.0 & 0.62 & 67\% \\
DP Lap. & LB-KPM & 150 & 0.1 & 0.63 & 67\% & DP Pix. & LB-KPM & 150 & 40.0 & 0.62 & 68\% \\
DP Lap. & LBM & 100 & 10000.0 & 0.63 & 68\% & DP Pix. & AEH-KPM & 100 & 20.0 & 0.61 & 65\% \\
DP Lap. & LB-KPM & 150 & 10.0 & 0.62 & 69\% & DP Pix. & LB-KPM & 100 & 10.0 & 0.60 & 62\% \\
DP Lap. & LBM & 100 & 0.1 & 0.62 & 68\% & DP Pix. & LB-KPM & 100 & 20.0 & 0.58 & 53\% \\
DP Lap. & AEH-KPM & 150 & 1.0 & 0.62 & 68\% & DP Pix. & LB-KPM & 100 & 30.0 & 0.58 & 50\% \\
DP Lap. & AEH-KPM & 150 & 100.0 & 0.62 & 69\% & DP Pix. & KPM & 200 & 10.0 & 0.58 & 51\% \\
DP Lap. & LB-KPM & 150 & 10000.0 & 0.62 & 68\% & DP Pix. & LB-KPM & 100 & 40.0 & 0.57 & 50\% \\
DP Lap. & LB-KPM & 150 & 100.0 & 0.62 & 68\% & DP Pix. & SH-KPM & 150 & 10.0 & 0.55 & 43\% \\
DP Lap. & AEH-KPM & 150 & 10000.0 & 0.62 & 68\% & DP Pix. & KPM & 200 & 40.0 & 0.54 & 43\% \\
DP Lap. & LB-KPM & 150 & 1.0 & 0.62 & 68\% & DP Pix. & SH-KPM & 150 & 30.0 & 0.53 & 36\% \\
DP Lap. & LB-KPM & 150 & 100000.0 & 0.62 & 68\% & DP Pix. & KPM & 200 & 30.0 & 0.53 & 40\% \\
DP Lap. & AEH-KPM & 150 & 1000.0 & 0.62 & 68\% & DP Pix. & SH-KPM & 150 & 20.0 & 0.53 & 37\% \\
DP Lap. & LB-KPM & 150 & 1000.0 & 0.62 & 68\% & DP Pix. & SH-KPM & 150 & 40.0 & 0.52 & 37\% \\
DP Lap. & AEH-KPM & 100 & 1.0 & 0.61 & 65\% & DP Pix. & AEH-KPM & 50 & 10.0 & 0.51 & 29\% \\
DP Lap. & AEH-KPM & 100 & 100.0 & 0.61 & 62\% & DP Pix. & KPM & 200 & 20.0 & 0.50 & 36\% \\
DP Lap. & LBM & 100 & 1000000.0 & 0.60 & 63\% & DP Pix. & LB-KPM & 50 & 10.0 & 0.50 & 19\% \\
DP Lap. & AEH-KPM & 100 & 10.0 & 0.60 & 64\% & DP Pix. & AEH-KPM & 50 & 20.0 & 0.50 & 22\% \\
DP Lap. & AEH-KPM & 100 & 1000.0 & 0.60 & 63\% & DP Pix. & SH-KPM & 100 & 20.0 & 0.49 & 15\% \\
DP Lap. & AEH-KPM & 100 & 10000.0 & 0.60 & 63\% & DP Pix. & SH-KPM & 100 & 10.0 & 0.49 & 17\% \\
DP Lap. & AEH-KPM & 100 & 0.1 & 0.60 & 65\% & DP Pix. & LB-KPM & 50 & 20.0 & 0.48 & 14\% \\
DP Lap. & AEH-KPM & 100 & 100000.0 & 0.57 & 54\% & DP Pix. & SH-KPM & 100 & 40.0 & 0.48 & 12\% \\
DP Lap. & LB-KPM & 100 & 10000.0 & 0.57 & 53\% & DP Pix. & KPM & 150 & 10.0 & 0.47 & 27\% \\
DP Lap. & LB-KPM & 100 & 100.0 & 0.57 & 52\% & DP Pix. & KPM & 50 & 40.0 & 0.47 & 8\% \\
DP Lap. & LB-KPM & 100 & 1000.0 & 0.57 & 50\% & DP Pix. & SH-KPM & 100 & 30.0 & 0.47 & 11\% \\
DP Lap. & LB-KPM & 100 & 10.0 & 0.57 & 50\% & DP Pix. & KPM & 150 & 40.0 & 0.47 & 19\% \\
DP Lap. & LB-KPM & 100 & 0.1 & 0.57 & 51\% & DP Pix. & KPM & 150 & 20.0 & 0.46 & 18\% \\
DP Lap. & LB-KPM & 100 & 1.0 & 0.56 & 52\% & DP Pix. & KPM & 150 & 30.0 & 0.46 & 19\% \\
DP Lap. & LB-KPM & 100 & 100000.0 & 0.56 & 47\% & DP Pix. & SH-KPM & 50 & 20.0 & 0.46 & 2\% \\
DP Lap. & SH-KPM & 150 & 10.0 & 0.53 & 32\% & DP Pix. & SH-KPM & 50 & 10.0 & 0.46 & 1\% \\
DP Lap. & SH-KPM & 150 & 100000.0 & 0.53 & 34\% & DP Pix. & KPM & 50 & 10.0 & 0.46 & 2\% \\
DP Lap. & SH-KPM & 150 & 100.0 & 0.52 & 33\% & DP Pix. & KPM & 50 & 30.0 & 0.46 & 9\% \\
DP Lap. & SH-KPM & 150 & 10000.0 & 0.52 & 31\% & DP Pix. & KPM & 50 & 20.0 & 0.45 & 0\% \\
DP Lap. & SH-KPM & 150 & 0.1 & 0.52 & 33\% & DP Pix. & KPM & 100 & 50.0 & 0.43 & 5\% \\
DP Lap. & SH-KPM & 150 & 1.0 & 0.52 & 31\% & DP Pix. & KPM & 100 & 10.0 & 0.42 & 9\% \\
DP Lap. & KPM & 200 & 100.0 & 0.52 & 33\% & DP Pix. & KPM & 100 & 30.0 & 0.41 & 4\% \\
DP Lap. & SH-KPM & 150 & 1000.0 & 0.51 & 33\% & DP Pix. & KPM & 100 & 40.0 & 0.41 & 4\% \\
DP Lap. & KPM & 200 & 0.1 & 0.51 & 35\% & DP Pix. & KPM & 100 & 20.0 & 0.40 & 5\% \\
DP Lap. & KPM & 200 & 10.0 & 0.51 & 37\% & & & & & & \\
\cline{7-12}
DP Lap. & KPM & 200 & 100000.0 & 0.51 & 37\% & Exponential & LBM & 100 & 200.0 & 0.57 & 59\% \\
DP Lap. & AEH-KPM & 50 & 100.0 & 0.51 & 21\% & Exponential & LBM & 100 & 150.0 & 0.56 & 55\% \\
DP Lap. & KPM & 200 & 10000.0 & 0.51 & 36\% & Exponential & LBM & 100 & 100.0 & 0.53 & 45\% \\
DP Lap. & KPM & 200 & 1.0 & 0.51 & 33\% & Exponential & LBM & 100 & 50.0 & 0.49 & 27\% \\
DP Lap. & KPM & 200 & 1000.0 & 0.50 & 36\% & Exponential & KPM & 100 & 100.0 & 0.40 & 5\% \\
DP Lap. & AEH-KPM & 150 & 1000000.0 & 0.50 & 25\% & Exponential & KPM & 100 & 150.0 & 0.40 & 5\% \\
DP Lap. & LB-KPM & 150 & 1000000.0 & 0.50 & 22\% & Exponential & KPM & 100 & 200.0 & 0.40 & 4\% \\
DP Lap. & AEH-KPM & 50 & 1.0 & 0.50 & 19\% & Exponential & KPM & 100 & 50.0 & 0.38 & 3\% \\
DP Lap. & AEH-KPM & 50 & 0.1 & 0.50 & 21\% & & & & & & \\
\cline{7-12}
DP Lap. & AEH-KPM & 50 & 10000.0 & 0.50 & 19\% & Laplace & LBM & 100 & 200.0 & 0.61 & 64\% \\
DP Lap. & AEH-KPM & 50 & 10.0 & 0.49 & 23\% & Laplace & LBM & 100 & 150.0 & 0.58 & 61\% \\
DP Lap. & AEH-KPM & 50 & 1000.0 & 0.49 & 20\% & Laplace & LBM & 100 & 100.0 & 0.55 & 54\% \\
DP Lap. & LB-KPM & 50 & 100.0 & 0.49 & 10\% & Laplace & LBM & 100 & 50.0 & 0.50 & 34\% \\
DP Lap. & LB-KPM & 50 & 10.0 & 0.49 & 12\% & Laplace & KPM & 100 & 200.0 & 0.39 & 5\% \\
DP Lap. & LB-KPM & 50 & 10000.0 & 0.49 & 13\% & Laplace & KPM & 100 & 50.0 & 0.38 & 1\% \\
DP Lap. & LB-KPM & 50 & 0.1 & 0.49 & 13\% & Laplace & KPM & 100 & 100.0 & 0.38 & 3\% \\
DP Lap. & LB-KPM & 50 & 1.0 & 0.49 & 13\% & Laplace & KPM & 100 & 150.0 & 0.36 & 2\% \\
DP Lap. & LB-KPM & 50 & 1000.0 & 0.48 & 14\% & & & & & & \\
\cline{7-12}
DP Lap. & SH-KPM & 100 & 1000.0 & 0.48 & 12\% & Normal & LBM & 100 & 200.0 & 0.59 & 61\% \\
DP Lap. & SH-KPM & 100 & 10.0 & 0.48 & 12\% & Normal & LBM & 100 & 150.0 & 0.57 & 55\% \\
DP Lap. & SH-KPM & 100 & 1.0 & 0.48 & 10\% & Normal & LBM & 100 & 100.0 & 0.54 & 49\% \\
DP Lap. & SH-KPM & 100 & 0.1 & 0.47 & 10\% & Normal & LBM & 100 & 50.0 & 0.48 & 27\% \\
DP Lap. & SH-KPM & 100 & 100.0 & 0.47 & 11\% & Normal & KPM & 100 & 50.0 & 0.38 & 4\% \\
DP Lap. & KPM & 50 & 1.0 & 0.47 & 9\% & Normal & KPM & 100 & 100.0 & 0.38 & 3\% \\
DP Lap. & KPM & 150 & 100.0 & 0.47 & 19\% & Normal & KPM & 100 & 200.0 & 0.37 & 4\% \\
DP Lap. & SH-KPM & 50 & 10.0 & 0.47 & 1\% & Normal & KPM & 100 & 150.0 & 0.37 & 3\% \\
DP Lap. & SH-KPM & 100 & 10000.0 & 0.47 & 10\% & & & & & & \\
\cline{7-12}
DP Lap. & KPM & 150 & 1000.0 & 0.46 & 17\% & Uniform & LBM & 100 & 200.0 & 0.55 & 56\% \\
DP Lap. & SH-KPM & 100 & 100000.0 & 0.46 & 10\% & Uniform & LBM & 100 & 150.0 & 0.52 & 44\% \\
DP Lap. & KPM & 150 & 10.0 & 0.46 & 16\% & Uniform & LBM & 100 & 100.0 & 0.51 & 33\% \\
DP Lap. & AEH-KPM & 100 & 1000000.0 & 0.46 & 4\% & Uniform & LBM & 100 & 50.0 & 0.44 & 16\% \\
DP Lap. & SH-KPM & 50 & 1.0 & 0.46 & 3\% & Uniform & KPM & 100 & 200.0 & 0.38 & 5\% \\
DP Lap. & KPM & 150 & 10000.0 & 0.46 & 14\% & Uniform & KPM & 100 & 50.0 & 0.38 & 2\% \\
DP Lap. & AEH-KPM & 50 & 100000.0 & 0.46 & 7\% & Uniform & KPM & 100 & 150.0 & 0.38 & 2\% \\
DP Lap. & SH-KPM & 50 & 0.1 & 0.46 & 2\% & Uniform & KPM & 100 & 100.0 & 0.37 & 2\% \\
DP Lap. & SH-KPM & 50 & 10000.0 & 0.46 & 2\% &  & & & & &  \\
\end{longtable}
\twocolumn